\documentclass{aa} 
\usepackage{orcidlink}
\usepackage{natbib}
\usepackage{hyperref}
\hypersetup{colorlinks,linkcolor={cyan},citecolor={blue},urlcolor={purple}} 

\usepackage{graphicx}
\usepackage{txfonts}
\usepackage{epsfig}
\usepackage{placeins}
\usepackage{times}
\usepackage{rotating}
\usepackage{float}
\usepackage[caption = false]{subfig}
\usepackage{setspace}
\usepackage{lscape}
\usepackage{epstopdf}
\usepackage{tabularx}
\usepackage{caption}
\usepackage{booktabs}
\usepackage{multirow}
\usepackage{auto-pst-pdf}
\usepackage[toc,page]{appendix}
\usepackage{soul}
\usepackage{amsmath}

\definecolor{RedStrong}{rgb}{0.85, 0.1, 0.1}

\newcommand{\HIIphot}{\textsc{\HIIphot}}

\usepackage{titlesec}

\begin{document}

\title{Bulgeless Evolution And the Rise of Discs (BEARD)}

\subtitle{I. Physical drivers of the mass-size relation for Milky Way-like galaxies}

\authorrunning{C. Marrero-de la Rosa et al. }

\titlerunning{BEARD: I. Physical drivers of the mass-size relation for Milky Way-like galaxies }

\author{
C. Marrero-de la Rosa\inst{1,2} \and
J. Méndez-Abreu\inst{1,2} \and
A. de Lorenzo-Cáceres\inst{1,2} \and
S. Cardona-Barrero\inst{1,2} \and
J. Román\inst{3} \and
E. Arjona-Gálvez\inst{1,2} \and
M. Chamorro-Cazorla\inst{3,4} \and
E. M. Corsini\inst{5,6} \and
L. Costantin\inst{7} \and
V. Cuomo\inst{8} \and
C. Dalla Vecchia \inst{2,1} \and
A. Di Cintio\inst{1,2} \and
D. Fernández\inst{9,10,11} \and
D. Gasparri\inst{12} \and
E. Iodice \inst{13} \and
D. Mayya\inst{9} \and
L. Morelli\inst{12} \and
F. Pinna\inst{1,2} \and
A. Pizzella\inst{5,6} \and
D. Rosa-González\inst{9} \and
Y. Rosas-Guevara\inst{14,15} \and
O. Vega \inst{9} \and
S. Zarattini\inst{16}
}

\institute{Universidad de La Laguna, Dpto. de Astrofísica. Avda. Astrofísico Francisco Sánchez S/N. E-38200 S.C de La Laguna, Spain 
\and
Instituto de Astrofísica de Canarias. C/ Vía Láctea S/N, E-38205 San Cristóbal de La Laguna, Spain 
\and
Departamento de Física de la Tierra y Astrofísica, Universidad Complutense de Madrid, 28040 Madrid, Spain 
\and
Instituto de Física de Partículas y del Cosmos (IPARCOS), Facultad de Ciencias Físicas, Universidad Complutense de Madrid, E-28040 Madrid, Spain 
\and
Dipartimento di Fisica e Astronomia “G. Galilei”, Università di Padova, vicolo dell’Osservatorio 3, I-35122 Padova, Italy 
\and
INAF-Osservatorio Astronomico di Padova, vicolo dell’Osservatorio 2, I-35122 Padova, Italy 
\and
Centro de Astrobiología, CSIC-INTA, Ctra. de Ajalvir km 4, Torrejón de Ardoz, E-28850, Madrid, Spain 
\and
Departamento de Astronomía, Universidad de La Serena, Av. Raúl Bitrán 1305, La Serena, Chile 
\and
Instituto Nacional de Astrofísica, Óptica y Electrónica, Tonantzintla, 72840 Puebla, México 
\and
Planetarium La Enseñanza, Medellín, Antioquia, CP. 050022 Colombia 
\and
Canada–France–Hawaii Telescope, Kamuela, HI 96743, USA 
\and
Instituto de Astronomía y Ciencias Planetarias, Universidad de Atacama, Copayapu 485, Copiapó, Chile 
\and
INAF–Astronomical Observatory of Capodimonte, Salita Moiariello 16, 80131 Naples, Italy
\and
Donostia International Physics Centre (DIPC), Paseo Manuel de Lardizabal 4, 20018 Donostia-San Sebastian, Spain 
\and
Departamento de Física, Universidad de Córdoba, Campus Universitario de Rabanales, Ctra. N-IV Km. 396, E-14071 Córdoba, Spain
\and
Centro de Estudios de Física del Cosmos de Aragón (CEFCA), Plaza San Juan 1, 44001 Teruel, Spain 
}

\date{\today}

\abstract{
In the standard $\Lambda$ cold dark matter ($\Lambda$CDM) cosmology, galaxies grow through smooth accretion and hierarchical mergers. While this framework explains many large-scale structures, the existence of massive disc galaxies without prominent bulges—pure discs—remains a challenge. In this work, we investigate the physical origin of the scatter in the stellar mass–size relation of massive spiral galaxies, with a particular focus on bulgeless systems. Studying these systems is also key to understanding the evolutionary history of our own Galaxy, the Milky Way, which is known to host a low-mass bulge. We performed a structural analysis of 22 nearby bulgeless galaxies from the Bulgeless Evolution And the Rise of Discs (BEARD) survey. To minimise the scatter in the stellar mass–size relation, we adopted a proxy for the physically motivated definition for the galaxy size, based on the radius $R_1$, where the stellar mass surface density reaches $\Sigma_* = 1\,{M}_\odot\,\mathrm{pc}^{-2}$. For this purpose, we used deep $g$- and $r$-band imaging obtained with the 2.5 m Isaac Newton Telescope–Wide Field Camera. We derived surface brightness, colour, and stellar mass density radial profiles, which allowed us to obtain precise measurements of $R_1$. Point spread function (PSF) effects were corrected through star subtraction and wavelet deconvolution. BEARD bulgeless galaxies follow the tight stellar mass-$R_1$ relation defined in previous studies with a similar scatter ($\sim$ 0.1 dex). We also constructed the same relation using galaxies from the IllustrisTNG50 simulation. We find a morphological segregation contributing to the scatter of the relation, with bulgeless (BEARD-like analogues) and bulge-dominated galaxies defining the upper and lower envelope, respectively. We find that this morphological trend shown by the simulations is strongly correlated with the specific central stellar mass density, $\Sigma^{\mathrm{spec}}_{1,\mathrm{kpc}}$, defined as the stellar mass surface density enclosed within the central kiloparsec, normalised using the total galaxy mass. The observed discrepancy between observations and simulations can be attributed to the broader $\Sigma^{\mathrm{spec}}_{1,\mathrm{kpc}}$ distribution covered by our observed BEARD bulgeless galaxies. A deeper analysis of the physical driver of this morphological segregation reveals that the scatter in the mass--size relation is also related to the spatial configuration of merger events, rather than their frequency, with bulgeless systems tending to inhabit halos with a slightly higher spin. 
}


\maketitle

\section{Introduction}
\label{sec:intro}

The stellar mass–size relation of galaxies encodes key information about the processes that regulate galaxy growth across cosmic time. Observational studies have shown that this relation evolves markedly with redshift, such that galaxies of a given stellar mass appear systematically more compact at earlier epochs and grow in size towards the present day \citep{shen2003, trujillo2006, somerville2008, vanderWel2014, buitrago2024}. In particular, \citet{buitrago2024} report that Milky Way–like disc galaxies ($M_* \sim 5 \times 10^{10} M_\odot$) have increased their characteristic sizes by a factor of $\sim$2 over the last 8 Gyr, accompanied by a significant decrease in their stellar mass surface density at the disc edge. This evidence highlights the dramatic inside-out growth of discs, with an average radial expansion rate of $\sim$1.5 kpc Gyr$^{-1}$. This evolution is widely interpreted as the result of an interplay between in situ star formation, merger-driven mass assembly,
and the influence of the dark matter (DM) halos in which galaxies
reside.

Traditionally, galaxy sizes have been quantified using parameters such as the effective radius ($R_{\textrm{e}}$), defined as the radius enclosing half of the galaxy total light \citep{vaucouleurs1948, graham2005a}. While widely adopted, these metrics can be sensitive to the depth of the observations, assumed light profile, and effects of inclination, making them vulnerable to significant scatter, especially for galaxies with irregular or extended morphologies \citep{capaccioli1983, trujillo2006, mosleh2013, costantin2023}. More recently, alternative size definitions such as the $R_1$ radius—defined as the galactocentric distance where the stellar mass surface density reaches $1\,M_\odot\,\mathrm{pc}^{-2}$—have been introduced as empirical proxies for the physical edge of in situ star formation \citep{trujillo2020}. This threshold is motivated by the critical gas density required for star formation ($\Sigma_c \sim 3$--$10\,M_\odot\,\mathrm{pc}^{-2}$; \citealt{schaye2004}) combined with typical star formation efficiencies, which yield stellar mass surface densities of order $1$--$3$ $M_{\odot}/\textrm{pc}^{-2}$. For galaxies with stellar masses comparable to that of the Milky Way, this `star formation edge' is indeed close to $\Sigma_* \sim 1\,M_\odot\,\mathrm{pc}^{-2}$ \citep{martinezlombilla2019}. While the exact stellar mass surface density at the disc edge is expected to vary with both galaxy mass and morphology \citep{chamba2020, chamba2022, buitrago2024}, adopting $\Sigma_* = 1\,M_\odot\,\mathrm{pc}^{-2}$ provides a useful proxy for defining the size of the galaxy. This approach enables a practical and consistent comparison of galaxy sizes across different morphological types, particularly in the low surface brightness regime. Nevertheless, $R_1$ should not be regarded as a strict physical boundary, as its interpretation depends on the underlying morphology and is less meaningful for spheroid-dominated systems, whose growth is largely merger-driven.

Several recent studies have sought to investigate the physical origin of the mass–size relation and its scatter using cosmological simulations. For example, \citet{du2024} and \citet{ma2024} have highlighted the role of dark matter halo spin, concentration, and assembly history in regulating galaxy size. However, their analyses rely on conventional size metrics such as $R_{\textrm{e}}$, which may obscure or dilute underlying correlations due to intrinsic measurement uncertainties. More recently, \citet{arjona2025} tested the $R_1$ size definition in state-of-the-art cosmological simulations, showing that rotation-dominated systems, such as bulgeless galaxies, tend to occupy the upper envelope of the mass--size relation.

Theoretically, in the framework of hierarchical structure formation, the formation of galactic discs is tightly coupled to the properties of their host dark matter halos. Following the classical picture of \citet{fall1980}, and later developments by \citet{mo1998} and \citet{somerville2008}, the size of a disc is expected to scale with the virial radius and angular momentum of its halo, such that $r_{\mathrm{disc}} \propto \lambda r_{\mathrm{vir}}$, where $\lambda$ is the halo spin parameter. This provides a simple physical basis for connecting the dark and luminous components of disc galaxies. However, while this classical picture offers a compelling first-order framework, it does not fully capture the complex role of mergers in shaping both the luminous and dark matter components of galaxies \citep{governato2009,governato2010,brook2011,naab2017}. In particular, the existence of massive, bulgeless galaxies poses a challenge to the $\Lambda$ cold dark matter ($\Lambda$CDM) paradigm. In a hierarchical Universe, galaxies grow through frequent mergers and accretion, the survival—or even the formation—of dynamically cold, pure stellar discs without central bulges is not straightforward \citep{hopkins09}. Mergers, especially those with high mass ratios, are expected to trigger central mass concentration through gas inflows and violent relaxation, promoting bulge growth \citep{naab2014}. Yet, observational surveys find large, rotationally supported galaxies with no significant bulge, indicating that some systems may evolve through more quiescent or finely tuned assembly histories (\citealt{kautsch2009,barazza09,kormendy2010,costantin2020}).

Addressing this tension is one of the central goals of the Bulgeless Evolution And the Rise of Discs (BEARD, Méndez-Abreu et al. in prep.) survey. This project is designed to systematically characterise the structure, kinematics, stellar content, and formation pathways of a statistically significant sample of nearby bulgeless galaxies. BEARD is a multi-facility survey that combines deep broadband photometry, narrowband H$\alpha$ imaging, and long-slit and integral field unit (IFU) spectroscopy to constrain the merger history and stellar mass growth of massive bulgeless galaxies. Combined with the study of accurately matched samples from numerical simulations, the survey aims to uncover the physical conditions that allow bulgeless galaxies to survive in a hierarchical Universe.

In this work, we present a pioneering analysis of a sample of nearby bulgeless galaxies from the deep images obtained as part of the BEARD survey. Deep $g$ and $r$ band observations are used to derive surface brightness, colour, and stellar mass density radial profiles, enabling the measurement of precise structural parameters such as $R_1$ and central stellar mass density, $\Sigma_{1,\mathrm{kpc}}$, defined as the stellar mass surface density enclosed within the central kiloparsec. To place our observational results in a theoretical context, we also make use of simulated galaxies from the IllustrisTNG50 cosmological simulation \citep{nelson2019, pillepich2019}, enabling a direct comparison between observed and simulated structural properties. By applying the same analysis techniques to both observed and simulated datasets, we are able to perform a direct, one-to-one comparison of the structural parameters. We focus on the connection between the stellar structure and properties of the dark matter halo and merger history, exploring whether halo spin and accretion-driven processes can explain the diversity in disc extents at fixed stellar mass. This approach provides a robust framework to test whether current $\Lambda$CDM-based models can reproduce the existence and characteristics of bulgeless galaxies in the local Universe.

The paper is organised as follows. In Sect.~\ref{sec:sample}, we describe the sample selection and observational data used in this study. Sect.~\ref{sec:methods} outlines the data reduction procedures and the methodology employed to derive structural parameters, including the $R_1$ radius and central stellar mass density. In Sect.~\ref{sec:results}, we present the stellar mass–size relation for bulgeless galaxies and investigate its connection to halo properties and merger histories using both observations and simulations. Finally, our main conclusions are summarised in Sect.~\ref{sec:conclusions}.

Throughout this study, we assume a standard $\Lambda$CDM cosmology with $\Omega_{\mathrm{m}} = 0.3$, $\Omega_{\Lambda} = 0.7$, and a Hubble constant of $\textrm{H}_0 = 70\,\mathrm{km\,s^{-1}\,Mpc^{-1}}$. For the simulated galaxies from IllustrisTNG50, whose internal cosmology slightly differs from these values, all distances have been corrected to ensure consistency with the adopted cosmology. All magnitudes are given in the AB photometric system.

\section{Sample and observations}
\label{sec:sample}

The galaxies presented in this work were observed as part of the BEARD survey (Méndez-Abreu et al. in prep.), where the volume-limited BEARD parent sample consists of 75 spiral galaxies selected from the 13th Data Release of the Sloan Digital Sky Survey (SDSS-DR13) spectroscopic catalogue \citep{sdss2017} and HyperLEDA database \citep{makarov2014}. The sample includes galaxies with stellar mass $\log \left( M_{*}/M_{\odot} \right) > 10$ located within a distance of 40 Mpc. To select systems without prominent bulges, we applied a concentration cut of $\textrm{C} = {R}_{90}/{R}_{50} < 2.5$ \citep{strateva2001,graham2005b} to galaxies with low inclination ($i < 60^\circ$), where $R_{90}$ and $R_{50}$ are the radii enclosing 90\% and 50\% of the total Petrosian flux, respectively. A further refinement of the sample, based on multi-component photometric decompositions (Zarattini et al. in prep.), was used to define the fiducial sample of 54 bulgeless Milky Way-like galaxies of the BEARD parent sample. In these decompositions---which include separate components for bulge, disc, bar, and unresolved central sources---we imposed $B/T < 0.1$ (bulge-to-total light ratio) and $B/D < 0.08$ (bulge-to-disc ratio), ensuring a robust selection of galaxies with negligible bulges.

To robustly estimate the location of the $R_1$ radius \citep{trujillo2020}, information on the colour of the stellar population is necessary. For each galaxy, exposures were taken in both the $g$ and $r$ bands of the Sloan Digital Sky Survey (SDSS; \citealt{york2000}) system. The observations were obtained using the Wide Field Camera (WFC) mounted on the 2.5 m Isaac Newton Telescope (INT) at the Roque de los Muchachos Observatory, spanning three observing proposals between 2019 and 2021 (see Table~\ref{tab:int_observations} in Appendix~\ref{sec:observations}). In total, 33 nights were allocated to the survey. The WFC consists of four CCDs arranged in an L-shaped mosaic, providing a total field of view of approximately $34 \times 34$ arcmins, with a pixel scale of $0.33$ arcsecs/pixel. The data reach limiting depths in surface brightness ranging from $\mu_{\textrm{g}} = 27.7$–$30.6 \ \textrm{mag} \ \textrm{arcsec}^{-2}$ and $\mu_{\textrm{r}} = 28.3$–$30.2 \ \textrm{mag} \ \textrm{arcsec}^{-2}$. These measurements corresponded to $3\sigma$ fluctuations relative to the image background, using $10 \times 10$ arcsec$^{2}$ boxes, following surface brightness limit definition by \citet{roman2020}.

From the volume-limited BEARD parent sample, a total of 40 galaxies were observed with this deep imaging strategy. However, we discarded those whose analysis was significantly affected by contamination from bright foreground stars in their outskirts and/or by the presence of galactic cirri, as both effects compromise the reliability of the surface brightness profile measurements. After this filtering process, we retained a final working sample of 22 galaxies, which are listed in Table~\ref{tab:full_sample}. Of these, 17 belong to the fiducial sample of 54 galaxies defined in Zarattini et al. (in prep.), while the remaining 5 do not.

\section{Methodology} \label{sec:methods}

\begin{figure*}[t]
    \centering
    \includegraphics[width =0.7\linewidth]{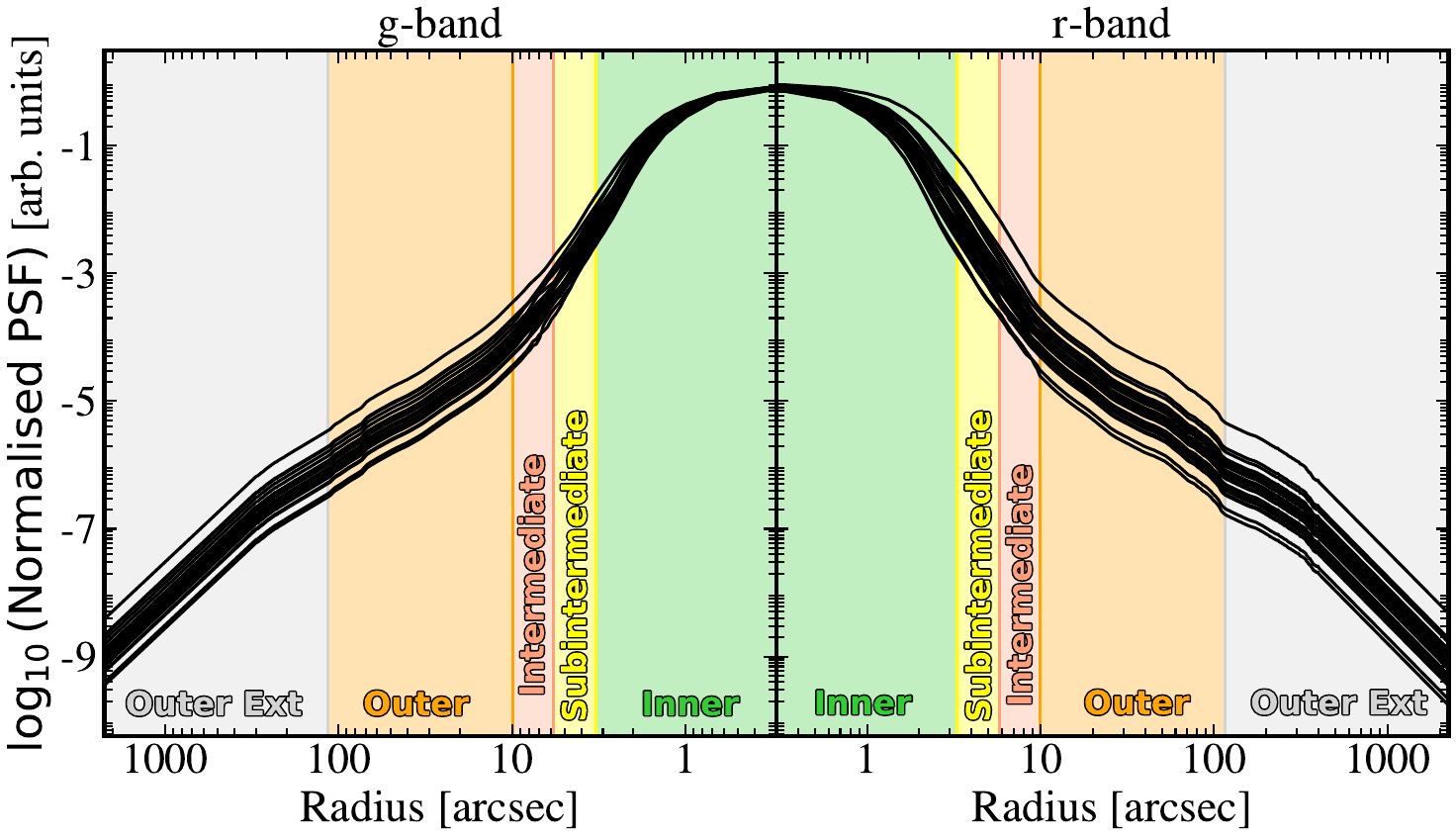}
    \caption{\footnotesize{Normalised radial profile of the PSFs for all fields observed in the $g$ and $r$ bands for the sample galaxies. The PSF total flux is scaled to unity. Vertical shaded regions indicate structural zones defined in the galaxy profile analysis (Inner, Subintermediate, Intermediate, Outer and Outer Ext) (see Sect.~\ref{sec:obtaining_psfs} for details). The y-axis represents normalised intensity in arbitrary units.}}
    \label{fig:all_psfs}
\end{figure*}

\subsection{The IllustrisTNG50 simulated sample}
\label{sec:tng50_sample}

For the simulated comparison, we adopted the same galaxy sample described in \citet{cardonaSubm}, extracted from the IllustrisTNG50-1 run \citep{pillepich2019, nelson2019}. This simulation follows the evolution of dark matter, gas, and stars in a $(51.7\,\mathrm{Mpc})^3$ volume, with a dark matter particle mass of $4.5 \times 10^{5}\,M_\odot$ and baryonic mass resolution of $8.5 \times 10^{4}\,M_\odot$. The assumed cosmology is consistent with \citet{planck2016}, but distances were corrected a posteriori to match the cosmology adopted in this work.

The high spatial resolution of IllustrisTNG50, with a gravitational softening length of 575 comoving pc until $z=1$ and fixed to 288 physical pc at lower redshifts, enables a detailed study of galaxy structure. We selected central galaxies with stellar masses $10^{10} < M_* < 5\times 10^{11}\,M_\odot$, applying an isolation criterion that excludes companions of comparable mass within 0.5 Mpc. This resulted in a sample of 537 simulated galaxies. For further details on the simulation setup and sample selection, we refer the reader to \citet{cardonaSubm}.

\subsection{Data reduction} 
\label{sec:reduction}
Image processing was carried out with the purpose of providing the highest reliability in the low surface brightness regime. For data reduction, we separated between image sets. Each set of images was formed by the blocks of nights indicated in Table \ref{tab:int_observations}, which are considered to have a uniform and similar flat-field between them, due to the temporal proximity of the observations.

Each set of images was bias-subtracted by standard procedures. For flat-fielding, we heavily masked the bias-subtracted images using \texttt{SExtractor} \citep{1996A&AS..117..393B} and \texttt{NoiseChisel} \citep{akhlaghi2015}. The coaddition of these masked images after normalisation provided a high-quality night-sky flat. Once this night sky flat was obtained, the images were flat-fielded and reduced. This procedure was performed independently for the $g$ and $r$ bands.

The process of obtaining the coadds---final stacked images created by combining multiple individual exposures to increase depth and signal-to-noise (S/N)---for each galaxy and each band was performed as follows. First, we computed astrometric solutions by using \texttt{Astrometry.net} \citep{2010AJ....139.1782L} to obtain an initial solution, which was then refined with \texttt{SCAMP} \citep{2006ASPC..351..112B} for greater accuracy. The next step was to produce a starting seed coadd, which served as the first step in an iterative process. This first coadd was made by stacking all the images of the field, performing a sky subtraction with a constant value. This preserved all the information in the low surface brightness regime, however, strong gradients were present in it. To provide a more efficient sky subtraction and to remove residual gradients while preserving the low surface brightness information we performed an iterative process. This process consisted of obtaining a mask from this coadd, and applying this mask to all individual exposures. Once the individual exposures were masked, we calculated the sky background using Zernike polynomials \citep{1934Phy.....1..689Z}. This sky background was subtracted from all individual exposures, which after photometric calibration referenced to SDSS, were combined to produce a new coadd. This process was repeated several times until a satisfactory coadd was obtained in which the residual gradients were minimised, while preserving the actual flux of the sources at low surface brightness. Typically, we used orders of Zernike polynomials between 2 and 4, depending on the intensity of the gradients present in the images. These gradients are variable between epochs and are especially intense in moderate moonshine conditions.

Similar coaddition processes have been carried out in the past \citep{2023A&A...677A.117S,2023A&A...679A.157R,2024A&A...689A.213P,2025A&A...694A.216J}, with highly efficient results. We thus made sure to avoid oversubtraction problems around sources, being particularly problematic the halos around galaxies of large apparent size, as the case of BEARD galaxies.

\subsection{Obtaining the extended PSF of the complete sample} \label{sec:obtaining_psfs}

It is well established that the PSF plays a critical role in the analysis of low surface brightness structures \citep{capaccioli1983, sandin2014, spavone2020, infante2020, sedighi2024}. In particular, the outskirts of galaxies may be significantly affected by scattered light originating from nearby stars as well as from the central regions of the galaxies themselves. Accurate PSF characterisation is therefore essential to ensure the reliability of the derived results. To this end, we followed a methodology consistent with previous works (\citealt{infante2020, sedighi2024}; Marrero-de la Rosa et al. in prep.). Rather than adopting a single, static PSF model, we constructed a tailored PSF for each observed field, accounting for variations in atmospheric seeing and instrumental response across the observing campaign. The PSF was constructed by stacking isolated stars from each field, grouped into concentric radial zones to accurately trace both the core and the extended wings. Unsaturated stars were used to define the inner PSF profile, while progressively brighter (and even saturated) stars were employed at larger radii, where their high flux allows us to recover the faint outer wings with improved S/N despite saturation in the central pixels. This multi-region strategy allowed us to recover the PSF structure out to large radii ($>1000$ arcsec), minimising background contamination and maximising radial coverage. As a result, a total of 22 distinct PSF profiles were derived in each filter ($g$ and $r$), one for each galaxy in the sample (Fig.~\ref{fig:all_psfs}).

To optimise both spatial resolution and S/N at all radii, the radial characterisation of the PSF was divided into five structural regions, following a procedure analogous to the one that will be presented in Marrero-de la Rosa et al. (in prep.). Each of these radial zones dominates the PSF profile in S/N within its respective range and boundaries between regions were chosen to ensure that adjacent profiles overlap at comparable S/N levels, guaranteeing smooth transitions across the full radial extent \citep{infante2020}. While stellar magnitudes are expressed in the \textit{Gaia} G-band \citep{gaia2016,gaia2023}, the PSFs were built from stars selected directly within each observed galaxy field, ensuring consistency with the specific observing conditions. The five regions, highlighted in Fig.~\ref{fig:all_psfs} with distinct colours, are defined as follows:

\begin{itemize}
    \item Inner (0--3.3 arcsec, highlighted in green): constructed using stars with magnitudes between 19 $<$ G $<$ 20 mag.
    \item Subintermediate (3.3--5.8 arcsec, in yellow): based on stars between 17 $<$ G $<$ 18 mag.
    \item Intermediate (5.8--9.9 arcsec, in salmon): based on stars between 15 $<$ G $<$ 16 mag.
    \item Outer (9.9--115.5 arcsec, in orange): modelled using the brightest field stars (G$<15$ mag), subdivided into five bins of magnitude to produce multiple independent stacks and improve robustness.
    \item Outer External ($>115.5$ arcsec, in grey): constructed using a single bright star (HD~8648) of G$=7.24$ mag located in the field of NGC~521. Such a bright source is essential to reliably trace the faintest outskirts of the PSF profile, where the signal from fainter stars becomes indistinguishable from the background noise.
\end{itemize}

Each radial region overlaps with the next within transition zones defined to ensure that both profiles share comparable S/N levels at the overlapping radius. This guaranteed smooth and continuous PSF profiles across the full dynamic range. The final PSFs resulted from the sequential stitching of these regions, calibrated to match flux at the overlap boundaries, and validated to avoid background oversubtraction or residual artefacts. Details of the corrections applied for scattered light from foreground stars, for self-scattered light from the galaxies themselves, and for masking contaminant sources and subtracting the local background can be found in Appendix~\ref{sec:app_stars}, Appendix~\ref{sec:psf_self_scatter} and Appendix~\ref{app:masking_data}, respectively.

\begin{figure*}[t]
    \centering
    \includegraphics[width =\linewidth]{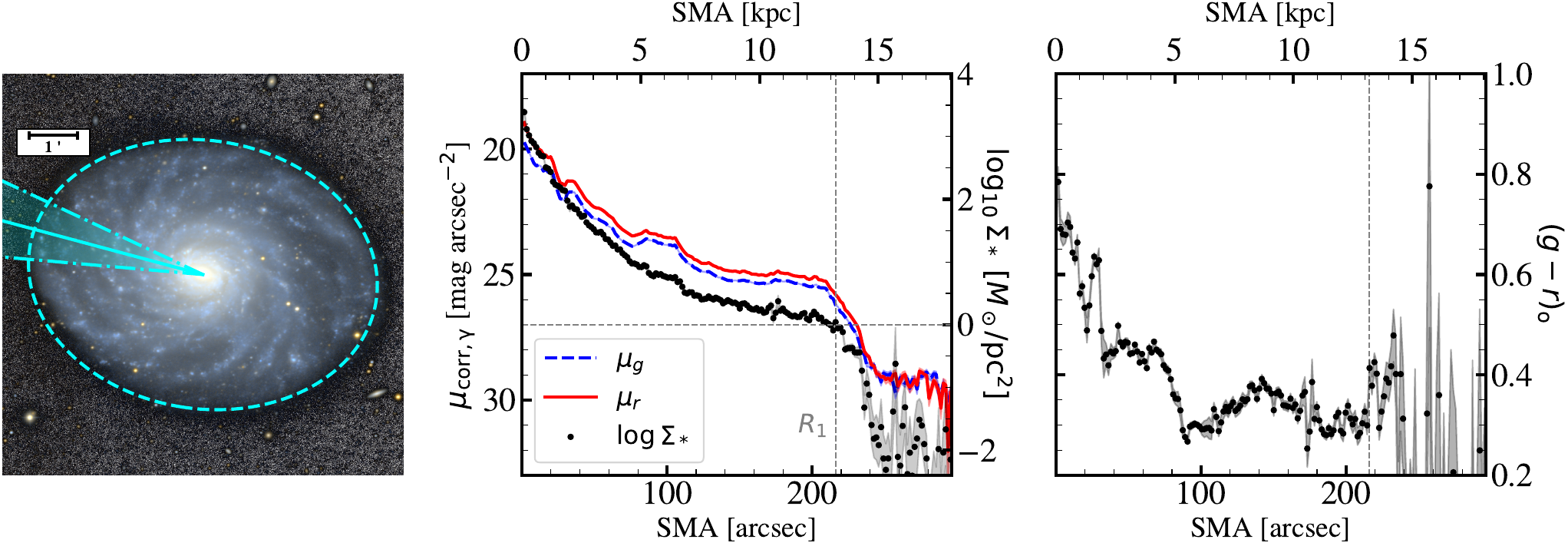}
    \caption{\footnotesize{
    Left panel: Colour image of NGC 3486. A 1 arcmin scale bar is shown, together with the area used to extract the profiles, highlighted as a wedge. The $R_1$ radius is shown as a cyan dashed-lined ellipse. 
    Middle panel: Surface brightness profiles in the $g$ (dashed blue) and $r$ (solid red) bands, along with the corresponding stellar mass surface density profile (black points). The vertical and horizontal dashed grey lines marks the $R_1$ radius and 1 $M_{\odot}~\textrm{pc}^{-2}$, respectively. Right panel: Colour profile $(g - r)_{\textrm{o}}$ corrected from extinction.}
    }
    
    \label{fig:profiles}
\end{figure*}

\subsection{Radial surface-brightness, colour, and stellar mass profiles}
\label{sec:SB}

Once all necessary image corrections have been applied, reliable surface brightness profiles can be derived. To ensure the robustness of these profiles, only the regions of the elliptical isophotes that intersect the wedge profiles—previously used for local background estimation—were considered (see Appendix~\ref{app:masking_data} for details). This procedure was carried out independently for both the $g$- and $r$-band images.

To place all galaxies on a common footing and mitigate projection effects, the surface brightness profiles were corrected for inclination following the prescription from \citet{trujillo2020}. This correction assumes that face-on profiles are systematically fainter than inclined ones due to geometric projection, and is implemented via a polynomial adjustment of the form

\begin{equation}
\Delta \mu = \sum_{j=0}^{4} \alpha_j \left(b/a\right)^j,
\end{equation}
\\

\noindent where $\Delta \mu$ is the inclination correction in mag $\textrm{arcsec}^{-2}$, $b/a$ is the observed axial ratio of the galaxy, and the coefficients $\alpha_j$ are determined empirically for a given $z_0/h$ ratio, with $z_0$ and $h$ being the vertical and radial scale lengths of the stellar disc, respectively. For this work, we adopted the coefficients corresponding to $z_0/h = 0.12$, as listed in Table~1 of \citet{trujillo2020}, which are representative of the late-type galaxies in our sample.

Additionally, all profiles were corrected for foreground Galactic extinction using the $A_{\gamma}$ values provided by the NASA NED Foreground Galactic Extinctions service \citep{schlafly2011}. These extinction corrections were independently applied for each photometric band and galaxy, ensuring a consistent comparison of intrinsic galaxy properties across the full sample. After applying both the inclination and extinction corrections, the final corrected surface brightness profiles used throughout this work are given by the following:

\begin{equation}
\mu_{\mathrm{corr},\gamma} = \mu_{\mathrm{obs},\gamma} - A_{\gamma} + \Delta \mu.
\end{equation}

This correction allows all galaxies to be consistently analysed as if observed face-on. From the inclination-corrected profiles, both the $(g - r)_{\textrm{o}}$ colour profiles and the stellar mass surface density profiles ($\Sigma_*$) were computed (Fig.~\ref{fig:profiles}).

To estimate the stellar mass surface density, we followed the approach described in \citet{bakos2008} and \citet{roediger2015}, using the relation

\begin{equation}
    \textrm{log}_{\textrm{10}} \Sigma_{*} =  \textrm{log}_{\textrm{10}} (M/L)_{\gamma} - 0.4\ (\mu_{\textrm{corr},\gamma}- m_{\odot,\gamma}^{\textrm{abs}}) + 8.629,
\end{equation}

\noindent where $\mu_{\mathrm{corr},\gamma}$ is the corrected surface brightness in a given $\gamma$ band, $m_{\odot,\gamma}^{\textrm{abs}}$ is the absolute magnitude of the Sun in that band, and $(M/L)_{\gamma}$ is the mass-to-light ratio calculated as

\begin{equation}
     \textrm{log}_{\textrm{10}} (M/L)_{\gamma} = a_{\gamma} + b_{\gamma} \times \textrm{colour}.
\end{equation}

In analogy with the methodology of \citet{chamba2022}, we adopted $\gamma = g$ as it provides the deepest photometric data. The colour term used was $(g - r)_{\textrm{o}}$, along with coefficients $a_g = -0.984$ and $b_g = 2.029$ taken from \citet{roediger2015}. The absolute magnitude of the Sun is $m_{\odot,g}^{\textrm{abs}}= 5.11$ from \citet{willmer2018}.

The $R_1$ radius was computed along the elliptical isophote where the stellar mass surface density reaches a value of $1 \ M_\odot \ \mathrm{pc}^{-2}$, with the galactocentric distance measured along the semi-major axis, i.e. not projected onto a wedge profile. This metric serves as an empirically motivated proxy for the extent of the stellar component in low surface brightness regimes, and it is particularly useful for comparing galaxies across a wide range of morphologies in both observations and simulations \citep{trujillo2020, arjona2025}. An example of the measurement of $R_1$, highlighted as a vertical grey line, can be seen in Fig.~\ref{fig:profiles}.

For the simulated galaxies from the IllustrisTNG50 project \citep{pillepich2019, nelson2019}, the $R_1$ radius was derived directly from the stellar mass density profiles. These were computed after centring each galaxy using the shrinking-sphere method of \citet{power2003}, reorienting the system so that the stellar disc is face-on with respect to its angular momentum vector, and then measuring the stellar mass density in logarithmically spaced radial bins. The $R_1$ radius was identified as the radius at which this profile reaches $1 \ M_\odot \ \mathrm{pc}^{-2}$, ensuring consistency with the observational definition while accounting for the three-dimensional nature of the simulation.

The total stellar mass of each galaxy, $M_*$, was obtained by integrating the stellar mass surface-density profile out to the radius where the $g$-band surface brightness profile reaches a limiting value of $29 \ \mathrm{mag \ arcsec^{-2}}$. This threshold was adopted to ensure consistency with the methodology employed in \citet{trujillo2020} and it represents a conservative limit, beyond which the photometric uncertainties become significant. In those particular cases where the surface brightness limit of $29 \ \mathrm{mag \ arcsec^{-2}}$ was not reached, the stellar mass was integrated up to the $R_1$ radius, which still encompasses the bulk of the galaxy stellar mass. Once the total mass had been computed using these criteria, the stellar mass–size relation could be derived.

\subsection{Environmental characterisation}
\label{sec:environment_method}

The potential impact of the surrounding environment on the scatter of the stellar mass–size relation was investigated by quantifying the local galaxy density for both the BEARD and IllustrisTNG50 samples. As a first step, we adopted the projected surface density to the fifth nearest neighbour ($\Sigma_{5}$), defined as $\Sigma_{5} = 5/({\pi r_{5}^{2}})$, where $r_{5}$ represents the projected distance to the fifth closest galaxy. This metric follows standard practice and has been employed in previous works (e.g. \citealt{aguerri2009}). Its specific application to the BEARD sample will be described in detail in the forthcoming survey paper (Méndez-Abreu et al. in prep.).

For the BEARD sample, $\Sigma_{5}$ was calculated using two complementary estimates: (i) a photometric version, based on SDSS imaging, and (ii) a spectroscopic version employing redshift information from SDSS–DR13 \citep{sdss2017} to minimise contamination from background or foreground sources. In both cases, neighbouring galaxies were required to have magnitudes within $\pm 2$~mag in the $r$ band relative to the host galaxy, ensuring that only systems of comparable luminosity were considered. The same magnitude selection was applied to the simulated galaxies when identifying neighbours within the IllustrisTNG50 volume.

The simulated galaxies were analysed using the same definition of $\Sigma_{5}$, applied to all central systems in the IllustrisTNG50 run. It is worth noting that these simulated galaxies were selected according to an isolation criterion that excludes companions of similar stellar mass within a projected radius of 0.5~Mpc. This selection ensures that the simulated sample resembles the typically isolated nature of the majority of BEARD galaxies, although it may bias the environmental characterisation towards lower-density regimes.

In addition to $\Sigma_{5}$, we also computed the instantaneous tidal radius ($r_{\mathrm{tidal,inst}}$) for the first five nearest neighbours of each central galaxy, defined as

\begin{equation}
r_{\textrm{tidal,inst}} = D \times \left[\frac{m_{\textrm{sat}}}{3M_{\textrm{host}}}\right]^{1/3},
\end{equation}

where $D$ is the instantaneous separation between galaxies and a circular orbit ($e_{\mathrm{orb}}=0$) is assumed. This formulation follows the classical tidal approximation (e.g. \citealt{king1962, johnston2002}), in which the ratio between the satellite and host enclosed masses determines the strength of the tidal field. Similar approaches have been applied in recent studies of diffuse tidal features and low-surface-brightness systems (e.g. \citealt{montes2021}). In this work, $r_{\mathrm{tidal,inst}}$ was computed using the stellar masses and separations of galaxies in the IllustrisTNG50 simulation, considering up to the five closest neighbours of each BEARD analogue.

\begin{figure*}[t]
\centering
\includegraphics[scale=0.65]{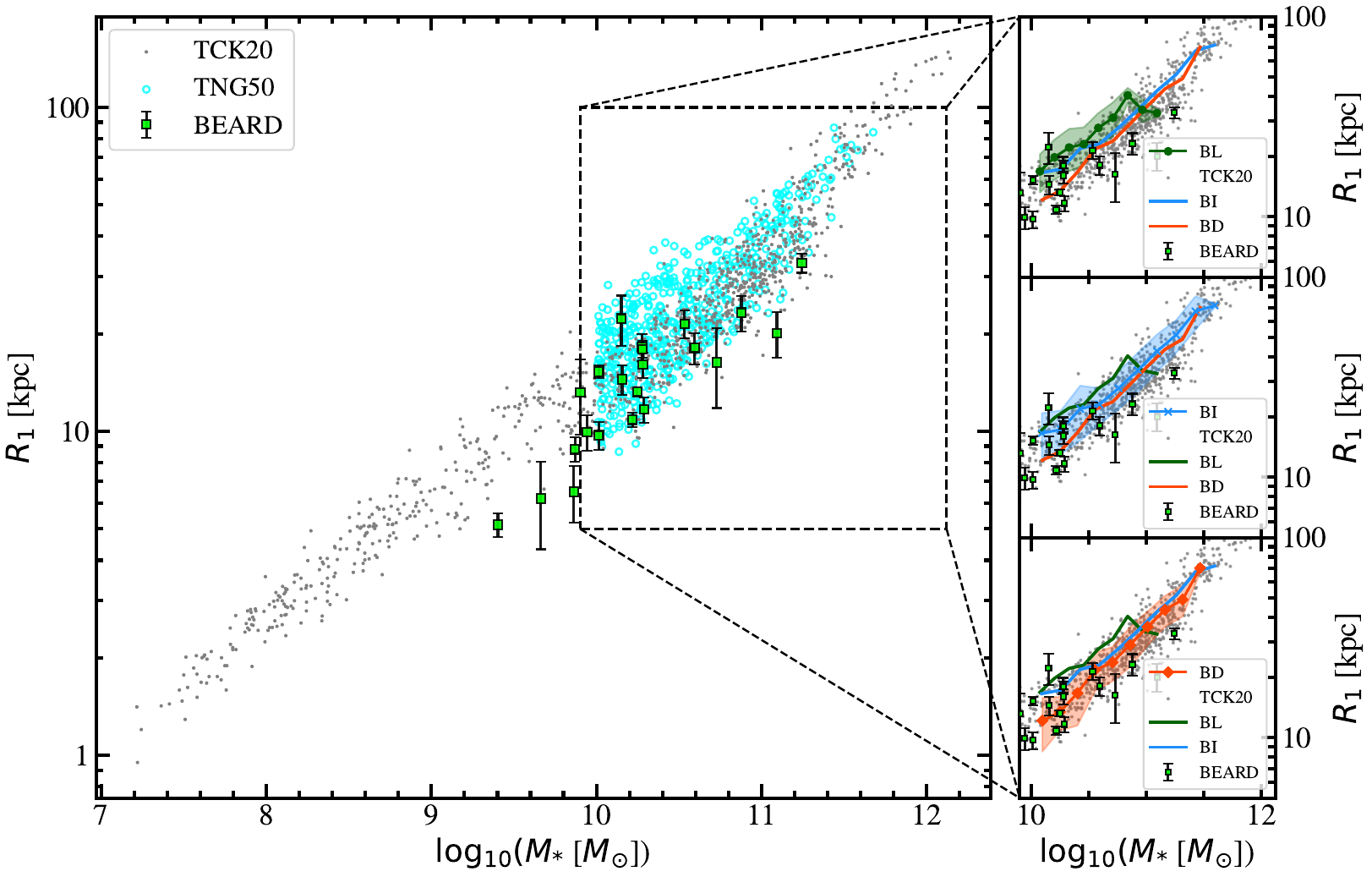}
\caption[\footnotesize{Mass-size}]{\footnotesize{Stellar mass–size relation defined by the $R_1$ radius. The main panel shows the global distribution of galaxies from different datasets: light-green squares show the BEARD sample of observed galaxies, grey points correspond to the observational sample from \citealt{trujillo2020} (TCK20), and cyan circles represent simulated galaxies from the IllustrisTNG50 run of the IllustrisTNG project \citep{cardonaSubm}. The right panels separate the IllustrisTNG50 sample into three categories based on their dynamically decomposed bulge-to-disc ratio: BL galaxies in green, BI galaxies in blue, and BD galaxies in orange. To compute the coloured lines for each morphological class, the stellar mass range was divided into 11 equally spaced bins; in each bin, the average $R_1$ value was calculated, and the shaded regions indicate the corresponding standard deviation. For comparison, the TCK20 (grey dots) and BEARD (light-green squares) samples are also shown in each panel.}}

\label{fig:mass_size}
\end{figure*}
\section{Results and discussion} \label{sec:results}
\label{sec:size}

In this section, we explore the stellar mass--size relation using $R_1$ as a physically motivated size indicator. Fig.~\ref{fig:mass_size} shows the stellar mass--size relation for the 22 galaxies from our BEARD sample (light-green squares). To place our results in a broader context and improve completeness, we include a supplementary dataset of 1005 galaxies compiled by \citet{trujillo2020}, shown as grey points. This ancillary sample comprises 279 ellipticals, 464 spirals, and 262 dwarfs, spanning a stellar mass range of $10^7 < M_* < 10^{12}\,M_\odot$ and redshifts between $0.01 < z < 0.1$. A detailed description of the sample selection and analysis can be found in the original publication. Although the BEARD and \citet{trujillo2020} samples differ in certain aspects of the data reduction and analysis, both followed broadly similar procedures to those outlined in Sect.~\ref{sec:reduction}. For the BEARD sample, the reductions were carried out using an updated version of the pipeline, and an additional wavelet-based deconvolution was applied to correct for self-scattered light from the galaxies themselves. In the overlapping mass range, the two datasets show good agreement.

Despite the fact that using $R_1$ as a size proxy yields a typical scatter in the mass–size relation that is approximately 2.5 times lower than that obtained with conventional size estimators based on $R_{\textrm{e}}$ \citep{trujillo2020}, there is a remaining scatter in the mass--size relation. In order to investigate it, we complement our analysis with a subsample of galaxies from the IllustrisTNG50 cosmological simulation \citep{cardonaSubm}. The aim is to provide a theoretical framework capable of interpreting the location and dispersion of the observed bulgeless galaxies within the mass--size plane (cyan circles, see Sect.~\ref{sec:tng50_sample}). 

Simulated galaxies are classified into three categories based on their bulge-to-disc (B/D) ratios: bulgeless (BL; also BEARD-Like, 134 galaxies), bulge-intermediate (BI, 269 galaxies), and bulge-dominated (BD, 134 galaxies). This classification is established by dividing the B/D distribution into quartiles, with thresholds at B/D$ = 0.057$ and B/D$ = 0.383$, where the bulge and disc components are dynamically identified through their orbital properties \citep{cardonaSubm}). This approach allows us to directly compare the structural diversity of galaxies across a continuous range of morphologies and to link the observed scatter with underlying physical parameters, such as halo properties and merger histories.

The BEARD sample is composed exclusively of bulgeless galaxies with B/D $\lesssim 0.08$. This allows for a direct comparison with the BL population in simulations and provides a stringent observational benchmark for the formation and structure of pure disc galaxies.

\subsection{The mass--size relation of Milky Way analogues}
\label{sec:masssize}

Within the stellar mass range covered by our sample of IllustrisTNG50 simulated spiral galaxies, a clear structural trend emerges in the mass--size relation (Fig.~\ref{fig:mass_size}). Galaxies classified as BL predominantly occupy the upper envelope of the relation, while BD systems tend to cluster on the lower side. Although this overall separation is evident, the dispersion bands shown in Fig.~\ref{fig:mass_size} partly overlap. This indicates that, at fixed stellar mass, BL galaxies are generally more extended than their BD counterparts, introducing an intrinsic scatter in the relation. Such a trend aligns well with the results from numerical simulations presented by \citet{arjona2025}, where the fastest rotating galaxies are preferentially located in the upper part of the mass--size diagram.

To quantify these trends, we performed linear fits to the mass--size relation for BEARD, BL, BI, and BD galaxies (Table~\ref{tab:mass_size_stats}; see Fig.~\ref{fig:fits_mass_size} in Appendix~\ref{sec:add_analysis}). For the IllustrisTNG50 sample, there are clear differences in both slope and intercept among the three morphological categories: BL galaxies define the upper envelope with shallower slopes and larger radii at fixed stellar mass, BD galaxies occupy the most compact regime with the steepest slope, and BI systems trace an intermediate sequence. When comparing BL galaxies in IllustrisTNG50 and BEARD, the fitted slopes are fully consistent, but a small offset in intercept is evident, with BEARD galaxies lying slightly below the simulated BL sequence. This offset is reinforced by the presence of a number of IllustrisTNG50 BL galaxies that deviate from the observed mass--size relation, lying outside the region populated by both the BEARD and \citet{trujillo2020} samples.

\renewcommand{\tablefoot}[2]{
  \par\noindent{Notes: #1}#2\par
}
\begin{table}[!ht]
\centering
\setlength{\tabcolsep}{3.0pt}
\caption{
 Statistical parameters of the $\log R_1$--$\log M_*$ relation.}
\begin{tabular}{lcccc}
\hline\hline
Sample & $\beta$ & $\sigma_{\mathrm{int}}$ [dex] & $r$ & $\alpha$ \\
\hline
BEARD & $0.38 \pm 0.05$ & $0.11 \pm 0.02$ & 0.846  & $-2.8 \pm 0.6$ \\
BL    & $0.39 \pm 0.03$ & $0.098 \pm 0.006$ & 0.708  & $-2.7 \pm 0.4$ \\
BI    & $0.43 \pm 0.02$ & $0.108 \pm 0.005$ & 0.847  & $-3.2 \pm 0.2$ \\
BD    & $0.55 \pm 0.02$ & $0.090 \pm 0.006$ & 0.919  & $-4.5 \pm 0.2$ \\
\hline
\end{tabular}
\vspace{0.2cm}
\tablefoot{ $\beta$ is the slope, $\sigma_{\mathrm{int}}$ the intrinsic scatter (in dex), $r$ the Pearson correlation coefficient, and $\alpha$ the intercept.
}

\label{tab:mass_size_stats}
\end{table}

\begin{figure*}[t]
    \centering
    \includegraphics[width =0.8\linewidth]{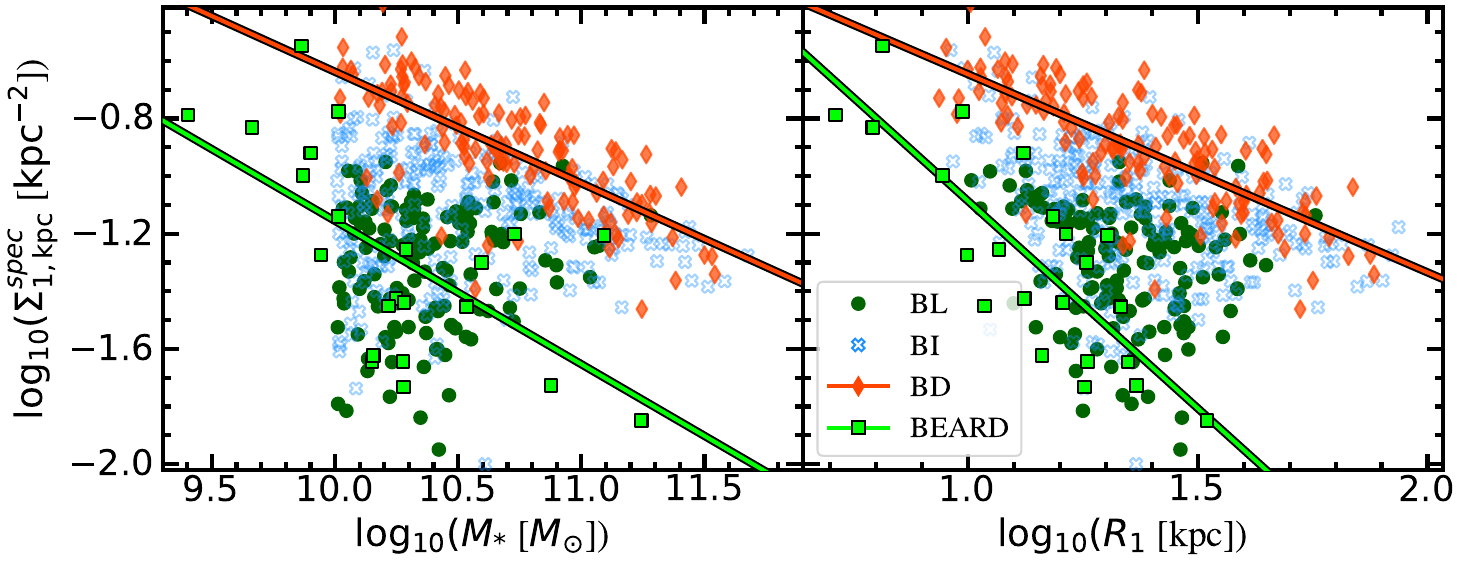}
    \caption[\footnotesize{Sigma1}]{\footnotesize{Relation between the specific central mass density $\Sigma_{1,\mathrm{kpc}}^{\mathrm{spec}}$ and the global properties of galaxies. Left panel: $\Sigma_{1,\mathrm{kpc}}^{\mathrm{spec}}$ as a function of stellar mass. Right panel: $\Sigma_{1,\mathrm{kpc}}^{\mathrm{spec}}$ versus $R_1$. In both panels, simulated galaxies from IllustrisTNG50 are colour-coded according to their bulge prominence, with BD galaxies (orange diamonds) occupying the upper regions of the relations, while BL systems (green circles) lie systematically below. BI galaxies (blue crosses) span a wide range in both $\Sigma_{1,\mathrm{kpc}}^{\mathrm{spec}}$ and size, bridging the two extremes. Observational data from the BEARD sample are shown as light-green squares, consistently occupying the region where BL galaxies are located.}}
    \label{fig:sigma1}
\end{figure*}

To further test the consistency between BEARD and IllustrisTNG50, we restricted both samples to the stellar mass range $10.0 < \log(M_*/M_\odot) < 10.5$ and generated $10^4$ random realisations of the BL subsample, each time drawing the same number of galaxies as in BEARD within this range. The resulting distribution of scatters is shown in Fig.~\ref{fig:scatter_comparison} in Appendix~\ref{sec:add_analysis}. The observed BEARD scatter ($\sigma_{R_1} = 0.11 \pm 0.02$) is fully consistent with the mean scatter of BL galaxies in IllustrisTNG50 ($0.11 \pm 0.02$). Kolmogorov--Smirnov (KS) tests \citep{stephens1974,fasano1987}, however, indicate that the distributions are statistically distinct: for the 1D comparison in $\log R_1$, we find $p = 5.2 \times 10^{-3}$, and for the 2D $(\log M_*, \log R_1)$ plane, $p = 1.1 \times 10^{-3}$. These values suggest that, while the level of scatter is comparable, the detailed distributions of the observed and simulated galaxies remain different.

Our BEARD galaxies do not preferentially lie on the upper envelope of the mass--size relation; rather, they span a broad range of $R_1$ values at fixed stellar mass, covering the full vertical extent of the relation. The error bars shown in Fig.~\ref{fig:mass_size} are computed by combining two main sources of uncertainty: first, the sensitivity of $R_1$ to the local background estimation in the wedge profiles---quantified by adding or subtracting the measured background uncertainty to the surface brightness profiles---and second, the uncertainty in stellar mass estimation. The latter follows the prescription of \citet{trujillo2020}, who compared their stellar mass estimates (derived using $g-r$ colour profiles and the $M/L$ relation of \citealt{roediger2015}) with independent values from the Portsmouth SED-fitting catalogue \citep{maraston2013}. For massive spiral galaxies analogous to those in our sample, they report an rms scatter of 0.24 dex, which we adopt here as a representative uncertainty in stellar mass. This choice is justified by the similarity of the adopted $M/L$ prescription and by the fact that both studies compute stellar masses by integrating the stellar mass surface density up to the limit of $29 \ \mathrm{mag \ arcsec^{-2}}$ in the $g$-band. Despite the intrinsic tightness of the mass--size relation when using $R_1$, these combined error bars do not fully account for the observed vertical scatter in most cases.

Therefore, in the following sections, we investigate potential physical drivers of this dispersion. First, we examine the connection between galaxy size and inner structure by measuring the stellar mass surface density within the central kiloparsec ($\Sigma_{1,\mathrm{kpc}}$) for both the BEARD observations and the IllustrisTNG50 simulations. We then turn to the simulated sample to explore how global halo properties---including total mass, concentration and spin parameter---may influence galaxy size. Finally, we analyse the assembly histories of the simulated galaxies, focusing on their merger rates and gas accretion patterns, to evaluate their impact on the structural diversity of present-day discs.

\subsection{Connecting the central and outer parts of the galaxy}
\label{sec:connect_central_outer}

A commonly used parameter to trace the central structure of galaxies is $\Sigma_{1,\mathrm{kpc}}$. This parameter has been shown to be a robust tracer of the internal structure of galaxies and their evolutionary stage \citep{cheung2012, vandoku2014, tacchella2015, lee2018, luo2020}. One of the key advantages of using $\Sigma_{1,\mathrm{kpc}}$ is that, unlike parameters that rely on bulge–disc decompositions or high-resolution imaging, it can be consistently measured across both observations and simulations using straightforward aperture photometry.

\begin{figure*}[t]
    \centering
    \includegraphics[width =0.8\linewidth]{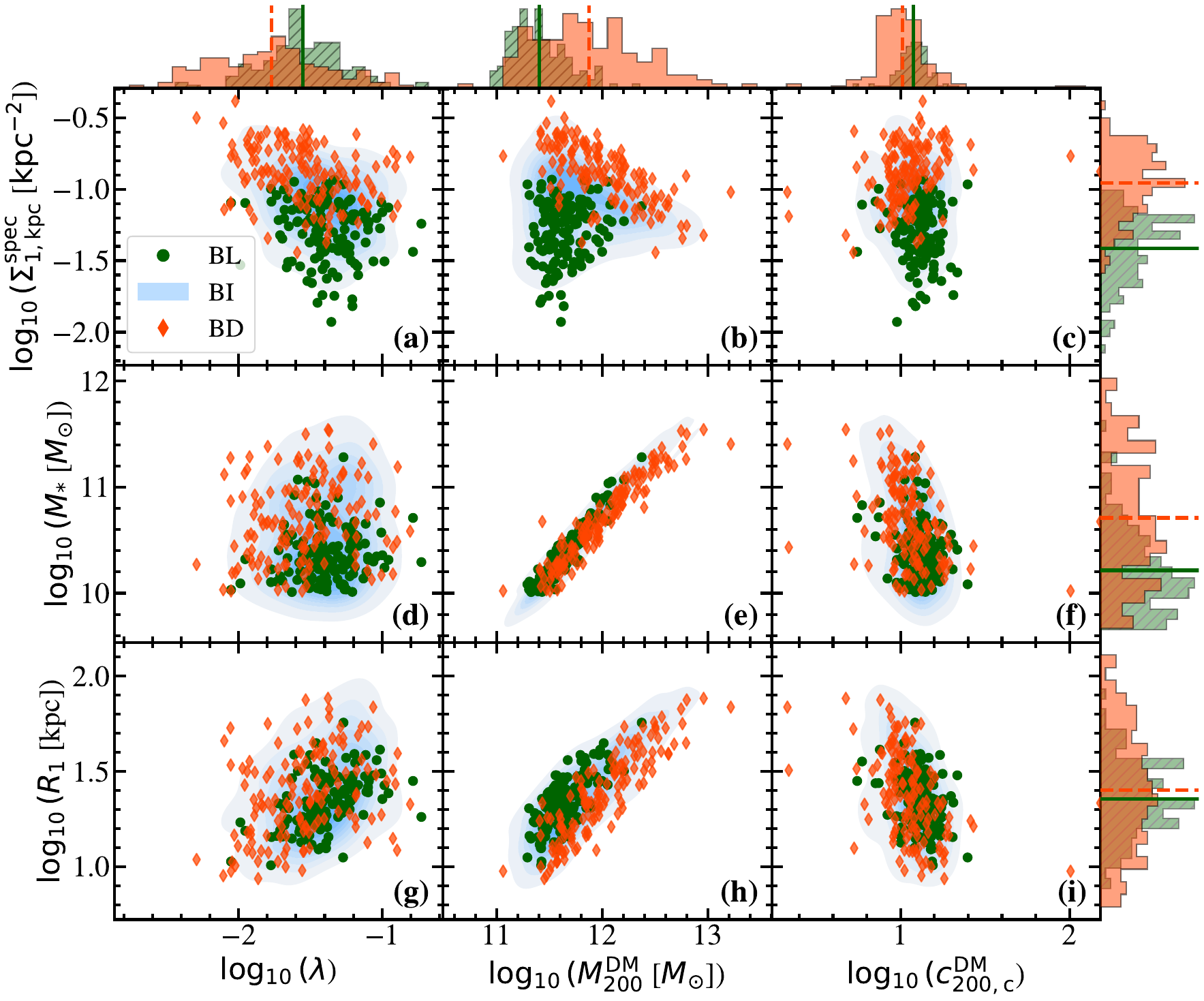}
    \caption[\footnotesize{Sigma1}]{\footnotesize{Relations between stellar and dark matter properties for the simulated galaxies from IllustrisTNG50. The figure is organised as a matrix, where each row corresponds to a stellar quantity (from top to bottom:  $\Sigma_{1,\textrm{kpc}}^{\mathrm{spec}}$, $M_{*}$ and $R_1$) and each column corresponds to a dark matter halo property (from left to right: $\lambda$, $M_{200}^{\mathrm{DM}}$, and $c_{200,\textrm{c}}^{\textrm{DM}}$). BL, BI, and BD galaxies are shown as green circles, blue isodensity contours, and orange diamonds, respectively. On the top and right panels, histograms of the corresponding quantities are shown for BL and BD galaxies, where vertical lines represent the mean of each distribution.}}
    
    \label{fig:baryonicVSdarkmatter}
\end{figure*}

However, to avoid the built-in correlation between total stellar mass and central mass density \citep{luo2020}, we normalise $\Sigma_{1,\mathrm{kpc}}$ by the total stellar mass of each galaxy, defining a specific central stellar mass density $\Sigma_{1,\mathrm{kpc}}^{\mathrm{spec}} \equiv \Sigma_{1,\mathrm{kpc}} / M_*$. This allows us to better isolate structural differences between galaxies independently of their stellar mass. Unlike our approach, which removes the stellar-mass dependence through this normalisation, \citet{luo2020} describe the correlation between $\Sigma_{1,\mathrm{kpc}}$ and $M_*$ using a quadratic polynomial fit in log–log space.

When connecting the inner structure of galaxies, traced by $\Sigma_{1,\mathrm{kpc}}^{\mathrm{spec}}$, with their global properties, we observe clear differences as a function of bulge prominence (Fig.~\ref{fig:sigma1}). BD galaxies follow a well-defined trend in both panels: in the left panel, $\Sigma_{1,\mathrm{kpc}}^{\mathrm{spec}}$ decreases with stellar mass with a slope of $-0.39$ and a Pearson correlation coefficient of $r=0.73$, while in the right panel the correlation with $R_1$ is similarly strong, with a slope of $-0.68$ and $r=0.77$. This highlights the tight link between central density and overall structure for bulge-dominated systems. By contrast, BL galaxies do not show a dominant correlation, with their central densities contributing mainly to the scatter of the mass–size relation. Interestingly, our BEARD galaxies exhibit a stronger correlation between $\Sigma_{1,\mathrm{kpc}}^{\mathrm{spec}}$ and $R_1$ (slope $-1.43$, $r=0.82$) than with stellar mass (slope $-0.50$, $r=0.62$), suggesting that for bulgeless discs the extent of the system is more closely connected to its central density. This result, however, may be affected by completeness biases in our observational sample. A more detailed exploration of these scenarios is presented in Sect.~\ref{sec:scatter_BL}.

\begin{figure*}[t]
    \centering
    \includegraphics[width =0.8\linewidth]{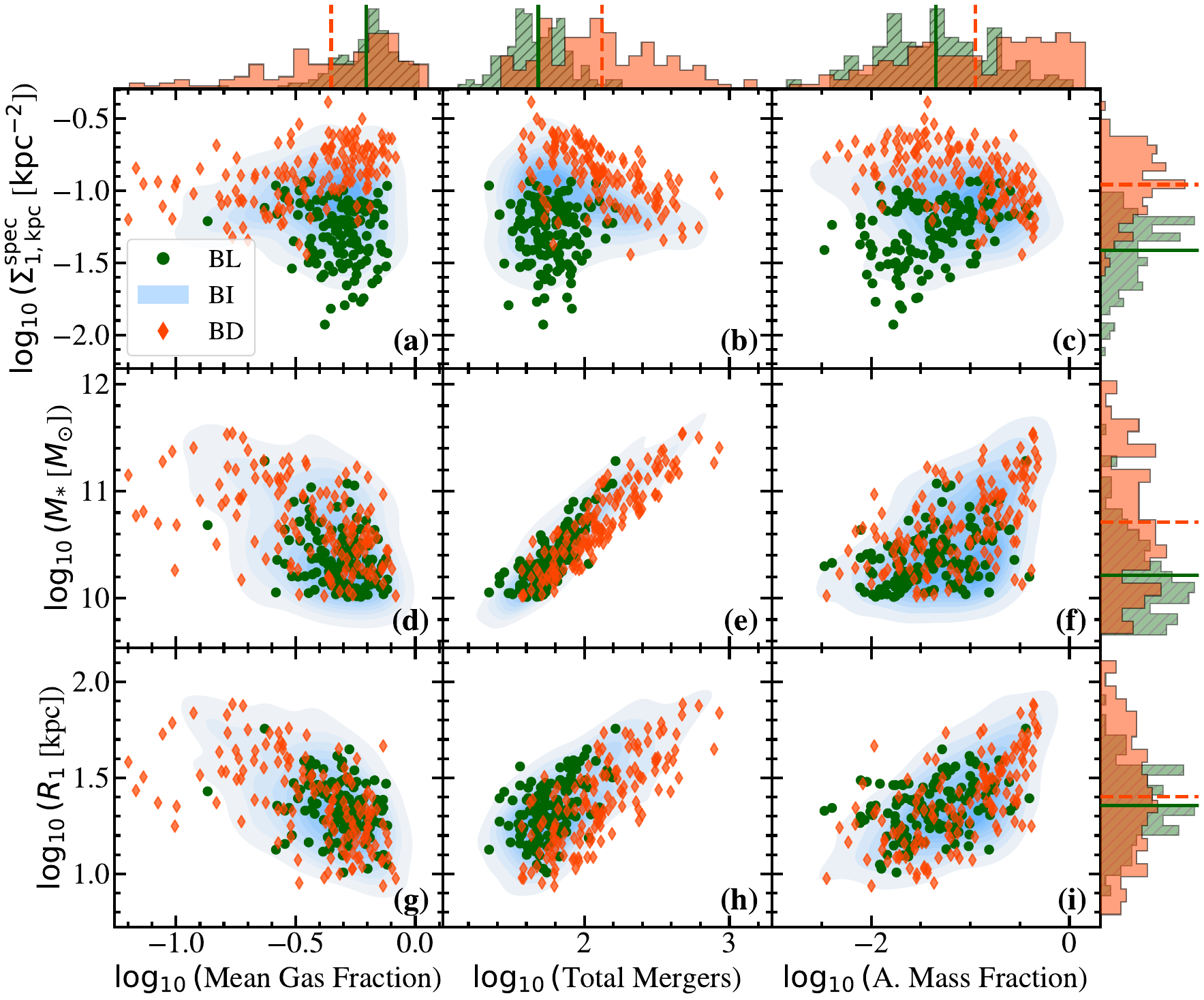}
    \caption[\footnotesize{Sigma1}]{\footnotesize{Relations between stellar and merger properties for the simulated galaxies from IllustrisTNG50. The figure is organised as a matrix, where each row corresponds to a stellar quantity (from top to bottom: $\Sigma_{1,\textrm{kpc}}^{\mathrm{spec}}$, $M_{*}$ and $R_1$) and each column corresponds to a merger property (from left to right: mean gas fraction, total mergers since $z=5$ and accreted mass fraction). BL, BI, and BD galaxies are shown as green circles, blue isodensity contours, and orange diamonds, respectively. On the top and right panels, histograms of the corresponding quantities are shown for BL and BD galaxies, where vertical lines represent the mean of each distribution.}}

    \label{fig:baryonicVSmergers}
\end{figure*}

\subsection{The impact of the dark matter halo}

To explore how the properties of dark matter halos influence the internal structure of spiral galaxies, we examine the relations between the halo mass ($M^{\rm DM}_{200}$), spin parameter ($\lambda$; defined as $\lambda = (J\,|E|^{1/2})/(G\,M^{5/2})$, where $J$, $E$, and $M$ are the total angular momentum, energy, and mass of the system, and $G$ is the Newtonian constant of gravitation, following \citealt{bullock2001}), concentration ($c_{200,\mathrm{c}}^{\mathrm{DM}}$), and key stellar structural properties such as $\Sigma_{1,\mathrm{kpc}}^{\mathrm{spec}}$ and $R_1$, using the simulated galaxies from IllustrisTNG50. Fig.~\ref{fig:baryonicVSdarkmatter} reveals multiple bimodal trends across this parameter space when these galaxies are separated by morphology.

A strong correlation is found between $\Sigma_{1,\mathrm{kpc}}^{\mathrm{spec}}$ and $M^{\rm DM}_{200}$ for BD galaxies, which follow a tight sequence (panel (b); Fig.~\ref{fig:baryonicVSdarkmatter}). BL and BI galaxies tend to deviate from this trend, populating the lower-density regime at fixed halo mass. This behaviour mirrors the correlation between $\Sigma_{1,\mathrm{kpc}}^{\mathrm{spec}}$ and $M_*$. When comparing $\Sigma_{1,\mathrm{kpc}}^{\mathrm{spec}}$ with $\lambda$ (panel (a); Fig.~\ref{fig:baryonicVSdarkmatter}), the negative correlation for BD galaxies becomes less clear due to larger scatter. However, a clear separation persists: at fixed spin, BD galaxies exhibit significantly higher central densities than BL systems. A similar pattern is observed at fixed $c_{200,\textrm{c}}^{\textrm{DM}}$, where BD galaxies again present systematically higher $\Sigma_{1,\mathrm{kpc}}^{\mathrm{spec}}$ values, suggesting that spin, concentration and mass jointly modulate the ability of halos to support dense stellar cores (panel (c); Fig.~\ref{fig:baryonicVSdarkmatter}).

Regarding stellar mass versus halo mass, a correlation with $M^{\rm DM}_{200}$ is evident (panel (e); Fig.~\ref{fig:baryonicVSdarkmatter}), with a small offset, that would result in a slight difference in the stellar-to-halo mass, similar to what we found in \citet{rosasguevaraSubm}. There is no robust trend for stellar mass versus concentration. But, analogously to the findings of \citet{proctor2024}, we also observe that halos with $\log_{10}(M^{\rm DM}_{200}/M_\odot) \gtrsim 12.5$ are predominantly populated by BD galaxies. Interestingly, the stellar mass versus spin relation shows that between $10.0 < \log_{10}(M_*/M_\odot) < 10.5$, BL galaxies tend to have higher spin values at fixed stellar mass.

We find a strong positive correlation between the $R_1$ radius and $M^{\rm DM}_{200}$, indicating that galaxy size increases with halo mass across the sample. Notably, at fixed $M^{\rm DM}_{200}$, BL galaxies tend to exhibit systematically larger $R_1$ values compared to their BD counterparts (panel (h); Fig.~\ref{fig:baryonicVSdarkmatter}). In contrast, neither the $R_1$–$c_{200,\textrm{c}}^{\textrm{DM}}$ nor the $M_*$–$c_{200,\textrm{c}}^{\textrm{DM}}$ relation display strong global trends, though morphological differences remain (panels (i) and (f) ; Fig.~\ref{fig:baryonicVSdarkmatter}).

These findings are consistent with both theoretical and observational works. Simulations by \citet{rodriguez2024} demonstrate that, at fixed stellar mass, disc-dominated galaxies tend to reside in lower-mass halos than spheroid-dominated ones. In addition, \citet{ma2024} report a clear link between the dynamical support of simulated disc galaxies and angular momentum of their host halos, highlighting the role of halo spin in sustaining rotationally supported structures. Observational studies support these trends through satellite counts, gravitational lensing, and dynamical tracers such as HI and globular clusters \citep[e.g.][]{wang2012, mandelbaum2016, posti2019}. Altogether, our analysis reinforces the view that both the mass and spin of dark matter halos—shaped in part by their assembly history—are fundamental in explaining the diversity of internal galaxy structures, especially the dichotomy between systems with and without prominent bulges.

\subsection{The role of merger history} \label{sec:roleofmergers}

To assess the impact of merger-driven processes on the internal structure of spiral galaxies, we analyse the relations between $\Sigma_{1,\mathrm{kpc}}^{\mathrm{spec}}$, $M_*$ and $R_1$ with respect to merger-related quantities: the mean `cold' (i.e. star-forming) gas fraction of all the objects that have merged with the galaxy, weighted by their maximum stellar masses, total number of mergers, both major (stellar mass ratio 1:4) and minor mergers since $z=5$, and accreted stellar mass fraction since $z=5$. All the information was taken from the publicly available IllustrisTNG50 catalogues \citep{rodriguez2017, eisert2023}. Fig.~\ref{fig:baryonicVSmergers} reveals several notable trends, many of which exhibit clear morphological bimodalities.

\begin{figure*}[t]
    \centering
    \includegraphics[width =0.85\linewidth]{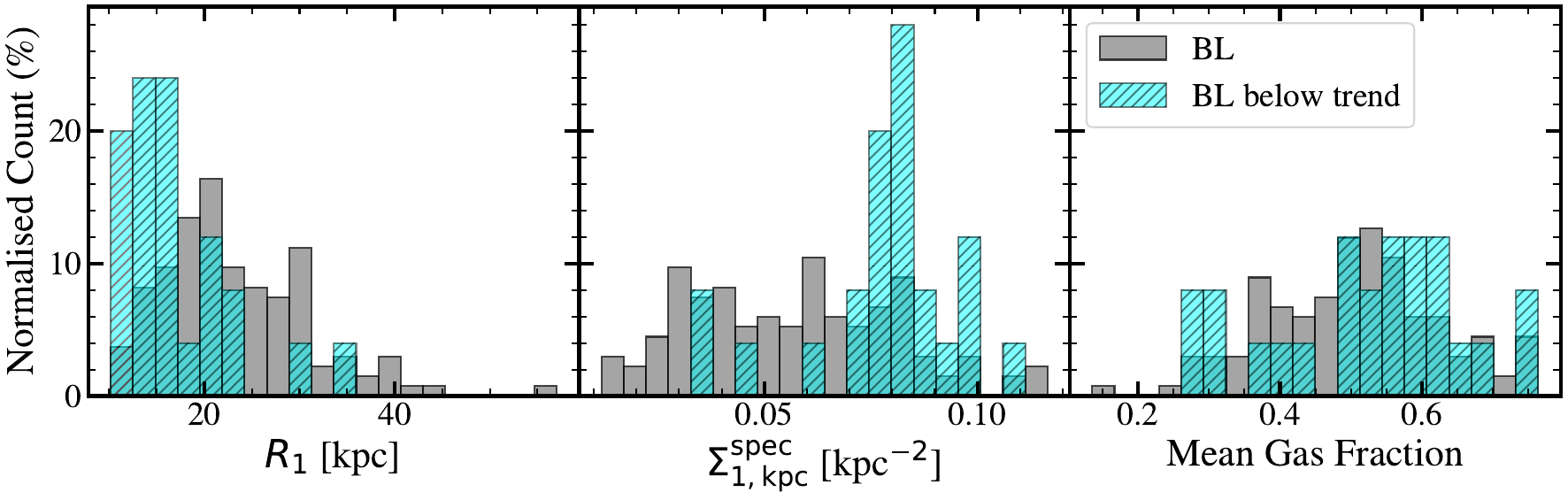}
    \caption[\footnotesize{BL_scatter}]{\footnotesize{Normalised distributions of three structural parameters for the simulated BL galaxies from IllustrisTNG50 (grey) and simulated BL galaxies that lie significantly below the average mass–size trend (hatched cyan). Left: $R_1$ radius,. Middle: central stellar mass density within 1 kpc, $\Sigma_{1,\mathrm{kpc}}^{\mathrm{spec}}$. Right: mean gas fraction since $z=5$.}}
    \label{fig:bl_scatter}
\end{figure*}

When comparing $\Sigma_{1,\textrm{kpc}}^{\rm spec}$ with the mean gas fraction (panel (a); Fig.~\ref{fig:baryonicVSmergers}), BD galaxies tend to exhibit lower gas fractions and higher central densities, while BL galaxies consistently occupy the lower-density, higher-gas-fraction regime. In particular, we do not find BL galaxies associated with mergers that had little cold gas, which is consistent with the idea that gas-poor (dry) mergers tend to produce spheroidal systems \citep{naab2009}. At higher gas fractions, both BD and BL morphologies coexist; however, BL systems remain systematically less centrally concentrated than their BD counterparts. This suggests that central compaction likely plays a key role in building dense inner regions, particularly in gas-poor merger events. The relation between $\Sigma_ {1,\textrm{kpc}}^{\rm spec}$ and the total number of mergers (panel (b); Fig.~\ref{fig:baryonicVSmergers}) echoes the behaviour seen with stellar and halo mass: BD galaxies follow a tight sequence where central density decreases with merger count, whereas BL systems show no clear trend and cluster towards lower $\Sigma_ {1,\textrm{kpc}}^{\rm spec}$ values. A similar dichotomy is seen when comparing $\Sigma_ {1,\textrm{kpc}}^{\rm spec}$ with the accreted stellar mass fraction (panel (c); Fig.~\ref{fig:baryonicVSmergers}): BD galaxies show a slight negative trend, while BL galaxies exhibit a positive correlation. This may indicate different evolutionary paths, with mergers playing contrasting roles depending on morphology.

In the $M_*$–mean gas fraction plane (panel (d); Fig.~\ref{fig:baryonicVSmergers}), no strong correlations are apparent, but a sharp morphological transition is evident: galaxies with $\log_{10}(\mathrm{Mean\ Gas\ Fraction}) \lesssim -0.6$ are exclusively BD, highlighting the role of gas depletion in bulge formation or preservation. When examining stellar mass versus total merger count (panel (e); Fig.~\ref{fig:baryonicVSmergers}), we find a clear positive correlation, with BD galaxies being more relevant at $\log_{10}(M_*/M_\odot) \gtrsim 11$ and total mergers $\gtrsim 100$. However, no significant trends emerge when stellar mass is plotted against the accreted mass fraction.

In terms of size, $R_1$ shows similar behaviour to $M_*$ when plotted against the mean gas fraction (panel (g); Fig.~\ref{fig:baryonicVSmergers}). However, the $R_1$–total mergers plane (panel (h); Fig.~\ref{fig:baryonicVSmergers}) reveals an important distinction: at fixed merger count, BL galaxies tend to be more extended than their BD counterparts. This supports the idea that merger-driven evolution contributes not only to central concentration but also to the scatter in the mass–size relation. A weak correlation is observed between $R_1$ and the accreted mass fraction (panel (i); Fig.~\ref{fig:baryonicVSmergers}), although no clear separation by morphology is found.

These results highlight the complex interplay between merger activity and galaxy structure. Our findings are broadly consistent with prior simulation-based studies that emphasise the importance of mergers in shaping both the morphology and compactness of galaxies. For instance, \citet{genel2015} and \citet{rodriguez2024} find that galaxies with richer merger histories develop more prominent bulges and denser cores. \citet{proctor2024} show that discs subjected to higher merger mass ratios become thicker and more extended, whereas more quiescent histories promote the survival of thin discs. Similarly, \citet{arianna2019} demonstrate that the orbital alignment of the last major merger can result in either high- or low-surface-brightness systems, producing two distinct morphologies at fixed stellar mass. Finally, \citet{jiang2025} report that giant BL galaxies at high redshift experience quieter merger histories compared to bulge-bearing systems. Furthermore, \citet{martig2009} suggest that mergers can also induce compaction in gas-rich systems, leading to high central densities. Our results support this view while extending it by showing how the role of mergers varies across morphology and may be a key driver of the observed scatter in structural scaling relations. The relation between merger history and dark matter halo properties has been further explored and is discussed in Appendix~\ref{sec:mergers_spin_mass}.

\begin{figure*}[t]
\centering
\begin{minipage}[t]{0.48\textwidth}
    \centering
    \includegraphics[width=0.95\linewidth]{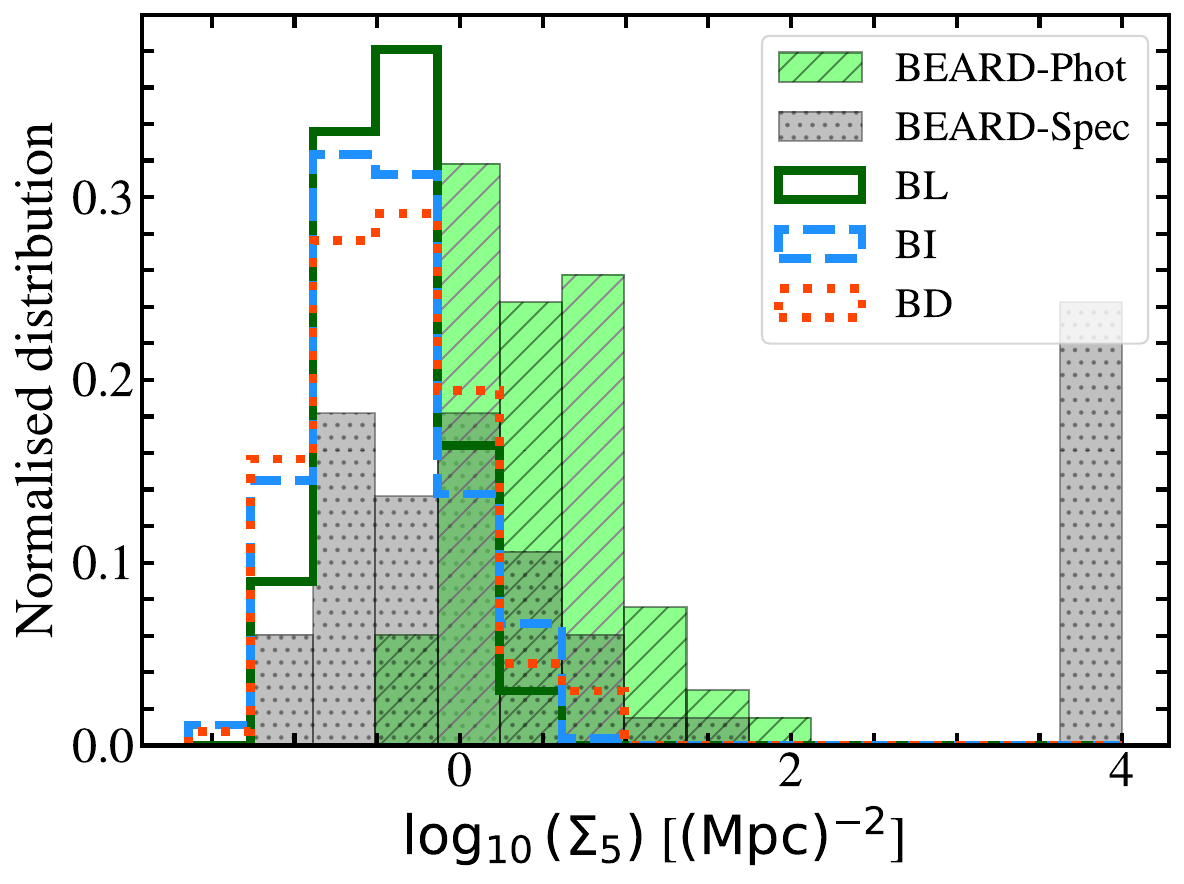}
    \caption{\footnotesize{
    Distribution of the projected environmental density parameter, $\Sigma_{5}$, for the simulated BL, BI, and BD galaxies compared with the BEARD-Phot and BEARD-Spec samples.}
    }
    \label{fig:sigma5_hist}
\end{minipage}
\hfill
\begin{minipage}[t]{0.48\textwidth}
    \centering
    \includegraphics[width=0.95\linewidth]{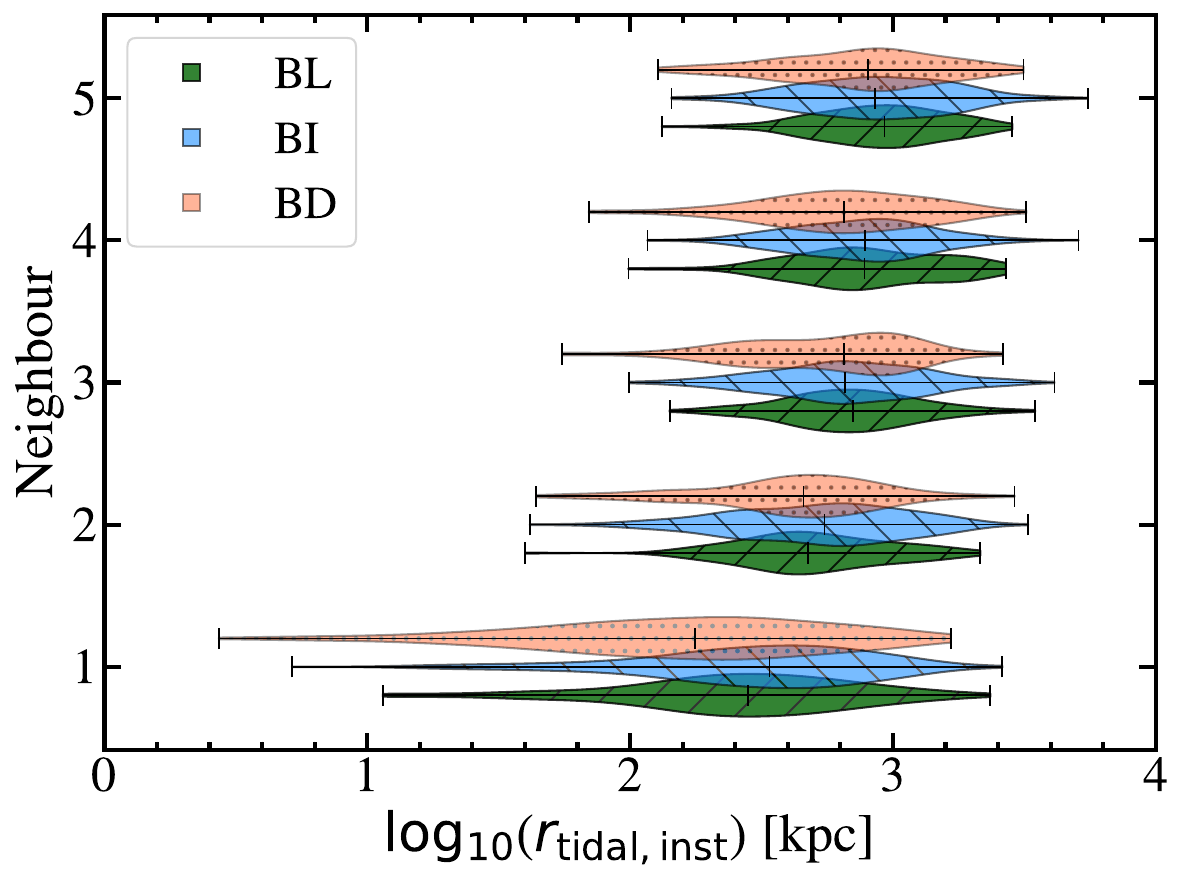}
    \caption{\footnotesize{
    Violin diagrams showing the distribution of the instantaneous tidal radius, $r_{\mathrm{tidal,inst}}$, for the first five nearest neighbours of each galaxy. The three morphological classes (BL, BI, BD) are displayed in different colours.}
    }
    \label{fig:rtide_violin}
\end{minipage}
\end{figure*}

\subsection{Intrinsic scatter within the bulgeless population}
\label{sec:scatter_BL}

While the stellar mass–size relation exhibits a bimodality between BD and BL galaxies, a level of intrinsic scatter is also present within the BL population itself. As shown in Fig.~\ref{fig:mass_size}, most of the simulated BL galaxies occupy the upper envelope of the relation, exhibiting large $R_1$ values at fixed $M_*$. This population is in good agreement with predictions from \citet{arjona2025}, where rotation-dominated galaxies are expected to exhibit more extended stellar discs due to their higher angular momentum content \citep{du2024}.

However, Fig.~\ref{fig:mass_size} also shows a significant number of BEARD BL galaxies falling below the simulated BL population. Besides any possible systematics in the comparison between observations and simulations, we also find a non-negligible fraction of BL systems in the IllustrisTNG50 sample that lie below the mean mass–size trend defined by BL galaxies (green line in Fig.~\ref{fig:mass_size}). This raises the question of what physical processes could account for such diversity within a morphologically homogeneous class. To investigate this, we selected a subpopulation of simulated BL galaxies from IllustrisTNG50 that systematically fall below the average $R_1$ value in bins of stellar mass by more than 1$\sigma$. A visual representation of this selected subsample can be found in Appendix~\ref{sec:app_bl_scatter}, where its location in the mass--size plane is shown more clearly.

The analysis of this subset reveals several key trends (Figs.~\ref{fig:bl_scatter}). Firstly, these `below-trend' BL galaxies exhibit, by definition, systematically smaller $R_1$ radii, as expected from the mass--size relation. Secondly, they show higher $\Sigma_{1,\mathrm{kpc}}^{\rm spec}$ values, similar to those found in BD systems, confirming their more compact nature. This configuration ---characterised by smaller $R_1$ and higher central densities--- mirrors the correlation between $\Sigma_{1,\mathrm{kpc}}^{\rm spec}$ and $R_1$ already identified in the BEARD galaxies (Sect.~\ref{sec:connect_central_outer}), reinforcing the view that central density is also a key factor behind their location in the mass--size plane. Finally, their mean gas fractions are slightly higher than those of typical BL galaxies, suggesting that these systems may be in an earlier or more gas-rich phase of disc evolution. This latter result is consistent with previous simulations where gas-rich mergers tend to create more compact and concentrated galaxies \citep{naab2009}.

These findings point towards a complex and heterogeneous evolutionary pathway for BL galaxies. The presence of compact, gas-rich discs with high central densities in the BL population may reflect different angular momentum acquisition histories, residual effects of minor mergers, or ongoing secular evolution. Disentangling these scenarios will require a more detailed dynamical characterisation of their stellar and gaseous components.

\subsection{Environmental trends}
\label{sec:environment_results}

Fig.~\ref{fig:sigma5_hist} shows the distribution of $\Sigma_{5}$ for the BEARD and IllustrisTNG50 samples. The simulated populations (BL, BI, and BD) display no clear morphological segregation in $\Sigma_{5}$, suggesting that large-scale environment does not play a dominant role in setting the structural differences among simulated galaxies. For the BEARD sample, the photometric estimate of $\Sigma_{5}$ tends to overpredict local densities, most likely due to projection effects, whereas the spectroscopic version yields values more consistent with those derived from the simulations. A small secondary peak is observed in the spectroscopic BEARD distribution, corresponding to galaxies located within the Virgo Cluster region, naturally reflecting their denser environments.

The distributions of the instantaneous tidal radius, $\log_{10}(r_{\mathrm{tidal,inst}})$, are presented in Fig.~\ref{fig:rtide_violin}. No significant differences are found among the three morphological categories (BL, BI, and BD) at any neighbour rank. The median values and overall dispersion of the distributions largely overlap, indicating that the tidal field strength experienced by these systems is similar across morphologies. Only for the first neighbour (i.e. the closest companion) is there a marginal trend for BD galaxies to reach slightly smaller $r_{\mathrm{tidal,inst}}$ values, which may indicate that their outer structures are more susceptible to the influence of nearby companions. However, this trend is weak and no significant differences are observed between the subsamples.

Overall, these results suggest that neither the large-scale environment, as traced by $\Sigma_{5}$, nor the local tidal field, as traced by $r_{\mathrm{tidal,inst}}$, exerts a dominant influence on the structural differences observed between BL, BI, and BD galaxies. The morphological segregation observed in the stellar mass–size relation therefore appears to be primarily governed by intrinsic factors such as the assembly history.

\section{Conclusions}
\label{sec:conclusions}

In this work, we have carried out a comprehensive structural analysis of a representative sample of nearby massive BL galaxies from the BEARD survey. Using very deep $g$- and $r$-bands imaging, we extracted surface brightness, colour, and stellar mass radial profiles for each system, applying rigorous corrections for PSF effects, scattered light contamination, and inclination. These corrections enabled a homogeneous and accurate measurement of key parameters, most notably the $R_1$ radius and the central stellar mass density $\Sigma_{1,\mathrm{kpc}}$. The inclusion of a matched galaxy sample from the IllustrisTNG50 simulation provides an important theoretical counterpart. We find that the simulations successfully reproduce the observed trends.

Simulated galaxies reveal that BL galaxies systematically occupy the upper envelope of the stellar mass–size relation, exhibiting significantly larger $R_1$ values at fixed $M_*$ compared to BD systems. This offset is accompanied by a markedly lower central stellar mass density, suggesting a fundamentally different evolutionary path for these galaxies. This trend is in agreement with the findings of \citet{arianna2019}, who showed that galaxies with lower central mass densities (i.e. lower surface brightness) develop more extended stellar distributions, resulting in larger effective radii. Although their simulations probed slightly lower stellar masses, at the low end of Milky Way–like systems, the qualitative consistency strengthens the interpretation of our results. Interestingly, while BD galaxies follow a tight sequence in the $\Sigma_{1,\mathrm{kpc}}$–$M_*$ plane, BL systems show much greater diversity in central density and disc extent. This trend is only marginally followed by our observed BEARD BL galaxies. Some BEARD galaxies are located in the upper part of the mass–size relation (Fig.~\ref{fig:mass_size}), as predicted by simulations, while others contribute to the scatter. Despite our limited number of observed galaxies, these differences might be attributed to variations in the central mass surface density of the galaxies.

Further analysis reveals that although we do not find strong global correlations between halo spin and merger-related parameters, BL galaxies tend to reside in halos with slightly higher spin. This may suggest that angular momentum still plays a role in enabling disc growth and preventing central classical bulge formation when merger configurations are favourable. However, the scatter in the mass–size relation does not appear to be primarily driven by the number of mergers experienced by each system, as both BL and BD galaxies show comparable merger counts since $z=5$. Instead, the key factor may lie in the specific configuration of those mergers---such as the relative orientation, halo spin, and orbital parameters---which could determine whether the disc structure is preserved or disrupted \citep{rosasguevaraSubm}. Moreover, our environmental analysis reveals no significant differences in local density or tidal field strength among the BL, BI, and BD populations, suggesting that the environment is unlikely to be the main driver of their structural diversity. However, this result may be partially influenced by the sample selection, particularly the isolation criteria applied to both the observed and simulated galaxies.

The BEARD survey provides a robust observational testbed for evaluating such scenarios and highlights the need to incorporate both dark matter halo properties and cosmological assembly paths when interpreting the structural diversity of disc galaxies in the local Universe.

\begin{acknowledgements}

The authors thank the anonymous referee for their thoughtful and constructive comments, which helped to improve the quality and clarity of this manuscript. The authors also sincerely thank Dr. Ignacio Trujillo for his valuable insights and constructive feedback throughout the development of this work. This project is possible thanks to financial support from the Spanish Ministry of Science and Innovation (MICINN) to the CoBEARD project (PID2021-128131NB-I00). JMA acknowledges the support of the Viera y Clavijo Senior programme funded by ACIISI and ULL and the support of the Agencia Estatal de Investigación del Ministerio de Ciencia e Innovación (MCIN/AEI/10.13039/501100011033) under grant nos. PID2021-128131NB-I00 and CNS2022-135482 and the European Regional Development Fund (ERDF) ‘A way of making Europe’ and the ‘NextGenerationEU/PRTR’. AdLC  acknowledges financial support from the Spanish Ministry of Science and Innovation (MICINN) through RYC2022-035838-I and PID2021-128131NB-I00 (CoBEARD project). Data for this paper has been obtained under the International Time Programme of the CCI (International Scientific Committee of the Observatorios de Canarias of the IAC) with the telescope operated on the island of La Palma by the operators in the Observatorio del Roque de los Muchachos. EAG acknowledges support from the Agencia Espacial de Investigación del Ministerio de Ciencia e Innovación (AEI-MICIN) and the European Social Fund (ESF+) through a FPI grant PRE2020-096361. ADC kindly thanks the Spanish Ministerio de Ciencia, Innovación y Universidades through grant CNS2023-144669, programa Consolidación Investigadora. MCC acknowledges the support of AC3, a project funded by the European Union's Horizon Europe Research and Innovation programme under grant agreement No 101093129. MCC acknowledges financial support from the Spanish Ministerio de Ciencia, Innovación y Universidades (MCIU) under the grant PID2021-123417OB-I00 and PID2022-138621NB-I00. EMC and AP acknowledge the support by the Italian Ministry for Education University and Research (MUR) grant PRIN 2022 2022383WFT "SUNRISE" (CUP C53D23000850006) and Padua University grants DOR 2022-2024. SZ acknowledges the financial support provided by the Governments of Spain and Aragón through their general budgets and the Fondo de Inversiones de Teruel, the Aragonese Government through the Research Group E16\_23R, and the Spanish Ministry of Science and Innovation and the European Union - NextGenerationEU through the Recovery and Resilience Facility project ICTS-MRR-2021-03-CEFCA. The project that gave rise to these results received the support of a fellowship from the "la Caixa" Foundation (ID 100010434). The fellowship code is LCF/BQ/PR24/12050015. LC acknowledges support from grants PID2022-139567NB-I00 and PIB2021-127718NB-I00 funded by the Spanish Ministry of Science and Innovation/State Agency of Research  MCIN/AEI/10.13039/501100011033 and by "ERDF" A way of making Europe. FP acknowledges support from the Horizon Europe research and innovation programme under the Maria Skłodowska-Curie grant "TraNSLate" No 101108180, and from the Agencia Estatal de Investigación del Ministerio de Ciencia e Innovación (MCIN/AEI/10.13039/501100011033) under grant (PID2021-128131NB-I00) and the European Regional Development Fund (ERDF) "A way of making Europe". VC acknowledges ANID (FONDECYT grant no. 11250723).
\end{acknowledgements}
\bibliography{references}

\begin{appendix}
\section{Observing Log and galaxy sample summary}\label{sec:observations}

This appendix provides additional information on the observational campaign and the final galaxy sample used in this work. Table~\ref{tab:int_observations} summarises the observations carried out with the Isaac Newton Telescope (INT) as part of three proposals between 2020 and 2021, including the number of nights allocated and those effectively observed. Table~\ref{tab:full_sample} lists the 22 galaxies that compose the final BEARD deep imaging sample. For each object, we report the J2000 coordinates, limiting surface brightness depths in the $g$ and $r$ bands, total stellar mass, and the $R_1$ radius.

\begin{table}[htbp]
\centering
\caption{Photometric observations with the INT.}
\label{tab:int_observations}
\begin{tabular}{cll}
\toprule
\toprule
Proposal ID & Dates (Number of nights) \\
\midrule
\multirow{4}{*}{ITP 2019-02} 
& 02.01.2020--04.01.2020 (3 nights) \\
& 27.01.2020--31.01.2020 (5 nights) \\
& 29.03.2020--02.04.2020 (5 nights)$^*$ \\
& 21.07.2020--23.07.2020 (3 nights) \\
& $^*$cancelled due to the Covid-19 pandemic \\
\midrule
\multirow{2}{*}{CAT47-INT7/19B} 
& 26.12.2019--01.01.2020 (7 nights) \\
& 12.01.2021--16.01.2021 (5 nights)$^*$ \\
& $^*$lost due to bad weather \\
\midrule
\multirow{4}{*}{ITP 2020-06} 
& 19.01.2021--22.01.2021 (4 nights) \\
& 02.02.2021--03.02.2021 (2 nights)$^*$ \\
& 29.04.2021--05.05.2021 (7 nights) \\
& 06.07.2021--08.07.2021 (3 nights) \\
& $^*$1 night lost due to bad weather \\
\midrule
\multicolumn{1}{r}{Total} & {44 nights / 33 nights observed} \\
\bottomrule
\bottomrule
\end{tabular}

\end{table}

\begin{table*}[htbp]
\centering
\caption{Observed properties and exposure times of the selected BEARD subsample.}
\label{tab:full_sample}
\setlength{\tabcolsep}{6pt}
\resizebox{\linewidth}{!}{
\begin{tabular}{lccccccc}
\toprule
\toprule
Name & RA (J2000) & DEC (J2000)  & $\textrm{SB}_{r}^{\rm lim}$ (mag $\textrm{arcsec}^{-2})$ & $\textrm{SB}_{g}^{\rm lim}$ (mag $\textrm{arcsec}^{-2})$ & $\mathrm{M}_{*}(M_{\odot})\times 10^{10}$ & $R_1$ (kpc)\\
(1) & (2) & (3) & (4) & (5) & (6) & (7)\\
\midrule
NGC 1087     & 02 46 24.9 & -00 29 46 &  30.2  & 30.5 &$0.74 \pm 0.01$ & $8.8 \pm 0.8$\\
NGC 1090     & 02 46 33.5 & -00 14 48 & 29.7  & 30.2 & $1.89 \pm 0.04$ & $18.2 \pm 1.7$\\
NGC 2543     & 08 12 58.4 & +36 15 03 &  30.2  & 30.6 &  $1.03 \pm 0.02$ & $15.3 \pm 0.7$\\
UGC 4375     & 08 23 11.3 & +22 39 50 &   29.6 &  29.9 & $1.90 \pm 0.04$  & $17.9 \pm 1.1$\\
$\textrm{NGC 2906}^*$    & 09 32 06.5 & +08 26 30 &  29.3 & 29.2& $5.4 \pm 0.4$ & $16.3 \pm 4.5$\\
$\textrm{NGC 3433}^*$     & 10 52 03.8 & +10 08 53 &  29.2 & 28.3& $3.41 \pm 0.05$  & $21.5 \pm 2.1$\\
NGC 3486     & 11 00 23.9 & +28 58 29 &   29.4 & 29.6 &  $0.79 \pm 0.04$  & $13.2 \pm 3.4$\\
NGC 3614     & 11 18 21.3 & +45 44 53 &  28.9   & 29.7& $1.4 \pm 0.1$  & $22.3 \pm 4.0$ \\
NGC 3683A    & 11 29 11.8 & +57 07 57 & 29.2  & 30.1&  $1.77 \pm 0.01$  &  $13.2 \pm 0.4$\\
$\textrm{NGC 3687}^*$    & 11 28 00.6 & +29 30 39 &  28.8  & 28.1&  $0.73 \pm 0.02$ & $6.5 \pm 1.3$\\
NGC 3756     & 11 36 48.5 & +54 17 37 & 29.3  & 29.3&   $1.43 \pm 0.03$ & $14.5 \pm 1.5$\\
NGC 3780     & 11 39 22.4 & +56 16 14 &  29.1  & 28.8&  $7.5 \pm 0.3$  & $23.2 \pm 2.9$\\
NGC 3810     & 11 40 58.8 & +11 28 16 &  29.2  & 28.9&   $1.03 \pm 0.02$ & $9.7 \pm 1.0$\\
NGC 3938     & 11 52 49.4 & +44 07 15 &   29.1 & 29.9&  $3.93 \pm 0.07$  &  $18.1 \pm 2.0$\\
NGC 4062     & 12 04 03.8 & +31 53 44 &  29.3  & 29.8&   $1.65 \pm 0.01$ & $10.9 \pm 0.5$\\
$\textrm{NGC 4405}^*$   & 12 26 07.1 & +16 10 52 &  28.7  & 29.7&  $0.46 \pm 0.03$  &$6.2 \pm 1.9$\\
NGC 4412     & 12 26 36.1 & +03 57 53 & 28.6  & 29.6&  $1.94 \pm 0.02$  &$11.7 \pm 1.0$\\
IC 3392      & 12 28 43.3 & +14 59 58 &  29.7  & 29.7&  $0.253 \pm 0.004$ & $5.1 \pm 0.4$\\
$\textrm{NGC 4540}^*$     & 12 34 50.9 & +15 33 06 &  29.8 &29.3&  $0.88 \pm 0.02$ & $9.9 \pm 1.2$\\
NGC 4814     & 12 55 21.9 & +58 20 39 &  28.3 &27.8&   $12.4 \pm 0.2$ & $20.1 \pm 3.2$\\
NGC 5660     & 14 29 49.8 & +49 37 22 &   29.8 &30.1&  $1.91 \pm 0.04$ & $16.1 \pm 1.5$\\
NGC 7606     & 23 19 04.8 & -08 29 06 & 29.4  & 29.9& $17.5 \pm 0.1$  & $33.1 \pm 2.2$\\
\bottomrule
\bottomrule
\end{tabular}}

\vspace{0.2cm}
\tablefoot{(1) Name of the galaxy. (2) Right ascension. (3) Declination. (4) and (5) Limiting depths in g and r bands. (6) Total stellar mass. (7) Estimation of $R_1 $ radius. Galaxies marked with an asterisk ($^*$) indicate those excluded from the fiducial sample of 54 galaxies from Zarattini et al. (in prep.).
}

\end{table*}

\renewcommand{\thefigure}{B.\arabic{figure}}
\setcounter{figure}{0} 
\begin{figure*}[h!]
    \centering
    \includegraphics[width=0.98\textwidth]{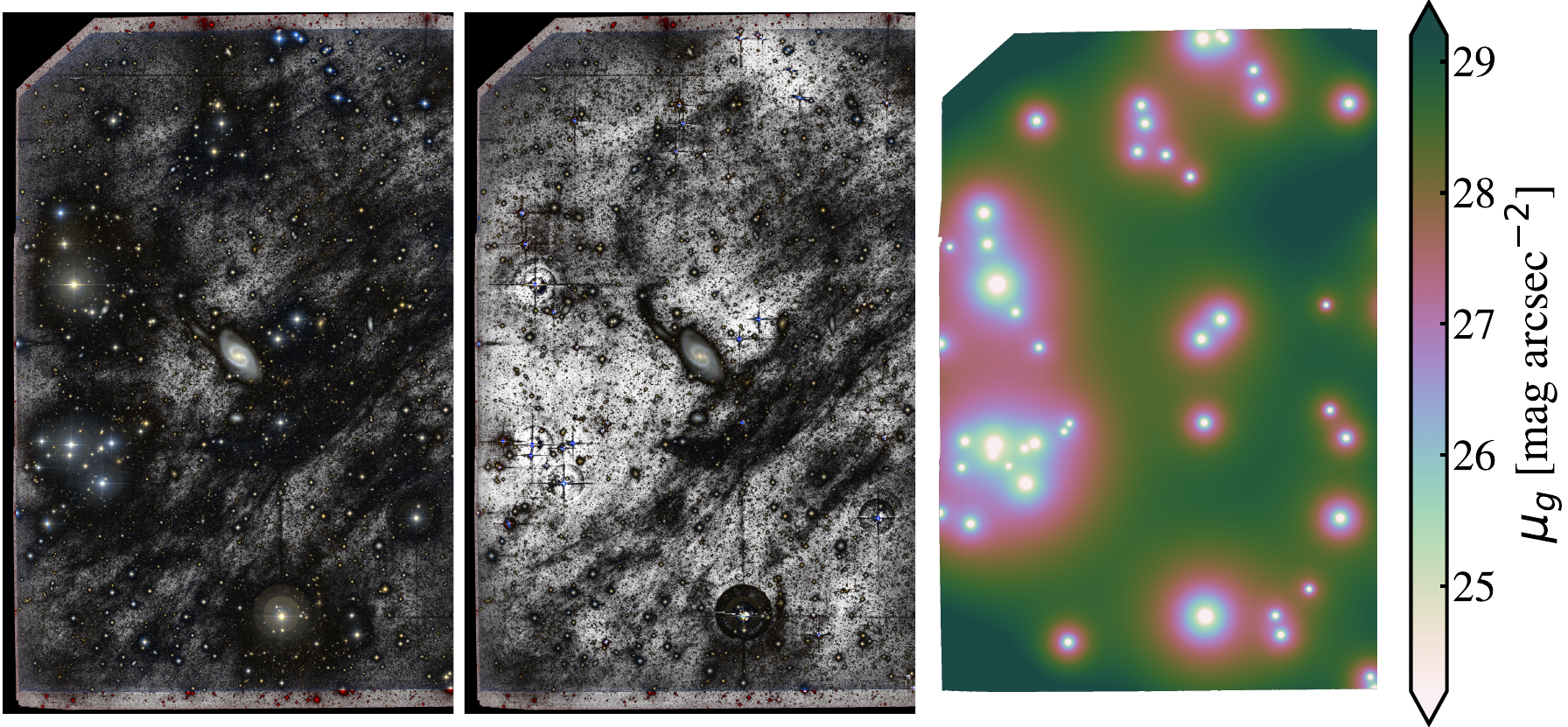}
    \caption{Example of stellar contamination and subtraction using the extended PSF. 
    Left: colour composite image of NGC2543, created using the SDSS $g$ and $r$ bands and the \texttt{astscript-color-faint-gray} algorithm described in \citet{infante2024}. 
    Middle: Same field after subtraction of bright stars. 
    Right: Scattered-light map, showing the extended halos of field stars.}
    \label{fig:star_subtraction}
\end{figure*}

\section{Subtraction of scattered light from foreground stars}
\label{sec:app_stars}

Following the characterisation of the PSF in both $g$ and $r$ bands, presented in Sect.~\ref{sec:obtaining_psfs}, we modelled and subtracted the scattered light field produced by foreground stars in each field of the survey. This process is essential, as previous studies showed that the extended halos of bright stars can act as a diffuse illumination screen with effective surface brightness levels of $\mu_{\textrm{V}} \sim$ 28--29 mag arcsec$^{-2}$, contaminating the detection and measurement of low surface brightness structures in galaxy outskirts \citep{slater2009, trujillofliri2016}. 

To ensure accurate PSF modelling, the first step involved constructing segmentation maps and masks for all detected sources in each field. These segmentation maps are pixel-based classifications that distinguish astronomical sources from the background, identifying which pixels belong to stars, galaxies, or artefacts. This process allowed us to mask contaminating sources and reliably isolate the stars used for building the PSF. This was achieved using \texttt{NoiseChisel} and \texttt{Segment} (\citealt{akhlaghi2015, akhlaghi2019}), two tools from the GNU Astronomy Utilities (Gnuastro) suite. These algorithms allow for robust detection of both compact and extended sources and assign unique labels to each detection, enabling their exclusion from the PSF analysis. The resulting masks are essential to prevent contamination from background galaxies or blended sources, particularly in the diffuse light regime. 

From the segmentation maps, a preliminary catalogue of detected sources was generated. This was then cross-matched with the \textit{Gaia} DR3 catalogue (\citealt{gaia2016,gaia2023}) using a 5 arcsec matching radius to account for potential positional discrepancies in saturated stars. Following the cross-match, an additional filter was applied to retain only those sources with either parallax or proper motion measurements exceeding three times their associated uncertainties, ensuring a clean stellar sample for PSF construction. 

Stars were selected based on their \textit{Gaia} G-band magnitudes, retaining only those brighter than 16 mag to ensure the removal of all bright stars whose extended scattered light may significantly contaminate the field. Each selected star was modelled individually using the extended PSF and subtracted from the image. The subtraction followed a hierarchical approach: brighter stars were addressed first, as their halos dominate the scattered light distribution. This ordering prevents contamination from bright sources during the modelling of nearby, fainter ones. Further technical details on the star selection and subtraction procedure can be found in Marrero-de la Rosa et al. (in prep.). 

Each star was modelled using a two-step process: first, the stellar centroid was fixed using \textit{Gaia} coordinates; second, the PSF model was scaled in flux to match the observed radial surface brightness profile of the star. To define a reliable normalisation region, we estimated the saturation radius $\textrm{R}_{\mathrm{sat}}$ as a function of stellar magnitude. This radius corresponds to the point beyond which the detector saturation no longer affects the observed profile. It was determined by analysing the first derivative of the radial surface brightness profile and identifying the radius at which the inner saturated region begins to deviate from the expected smooth decline of the PSF. Once $\textrm{R}_{\mathrm{sat}}$ was defined, the flux normalisation was performed over an annular region spanning $1.5\,\textrm{R}_{\mathrm{sat}}$ to $4\,\textrm{R}_{\mathrm{sat}}$, ensuring that the selected portion of the profile is both unsaturated and has a sufficient S/N for accurate scaling.

A scaling factor was computed as the $2\sigma$-clipped mean ratio between the observed and model PSF within this annulus. Once scaled, the PSF model was subtracted from the image. The final scattered light map was constructed by summing the contributions of all modelled stars. The surface brightness levels of the resulting maps were found to be consistent with typical values reported in the literature for diffuse stellar halos caused by scattered light, with effective levels of $\mu_{g,r} \sim 28$--29 mag arcsec$^{-2}$ (e.g. \citealt{slater2009, trujillofliri2016}). An illustrative example of this procedure, showing the original image, the star-subtracted field, and the corresponding scattered-light map, is presented in Fig.~\ref{fig:star_subtraction}. This procedure was applied uniformly across the full BEARD survey. To implement these procedures, we have developed dedicated Python tools, described in Marrero-de la Rosa et al. (in prep.), which will be publicly released through GitHub repositories. The first one, \texttt{LISAN} (Layered Intensity Spread and Analysis for Night-sky structures), is dedicated to the construction and calibration of extended PSFs over wide-field imaging datasets. The second, \texttt{MAHDI} (Mitigation Algorithm for Halo and Diffuse Illumination), performs scattered light subtraction by modelling and removing the halo of each star using the previously characterised PSF. Both codes will be publicly available via the following GitHub\protect\footnotemark\footnotetext{\url{https://github.com/CarlosMDLR/LISAN}\\ \url{https://github.com/CarlosMDLR/MAHDI}} repositories.

\section{Correction of self-scattered light via wavelet-based deconvolution}
\label{sec:psf_self_scatter}

In addition to the contamination from foreground stars, the extended wings of the PSF can scatter light from the galaxy itself into its outskirts. This effect can artificially flatten the surface brightness profiles and bias measurements of outer disc structures and stellar mass distributions.

To correct for this internal scattering, we applied a deconvolution algorithm\protect\footnotemark \footnotetext{\url{https://github.com/aasensio/Wavelet_deconvolution}} based on the isotropic undecimated wavelet transform (IUWT) \citep{starck1994,starck2010,carrillo2012}, as introduced in \citet{golini2025} and that will be explained in detail in Marrero-de la Rosa et al. (in prep.). This method leverages the multiscale decomposition of the image to separate smooth, extended structures (associated with PSF scattering) from compact and coherent galaxy features. The algorithm minimises a loss function that includes terms for the fidelity of the observed convolved image, smoothness of the reconstruction, and wavelet-based regularisation to suppress noise amplification.

\renewcommand{\thefigure}{C.\arabic{figure}}
\setcounter{figure}{0} 
\begin{figure}[tp]
    \centering
    \includegraphics[width =\linewidth]{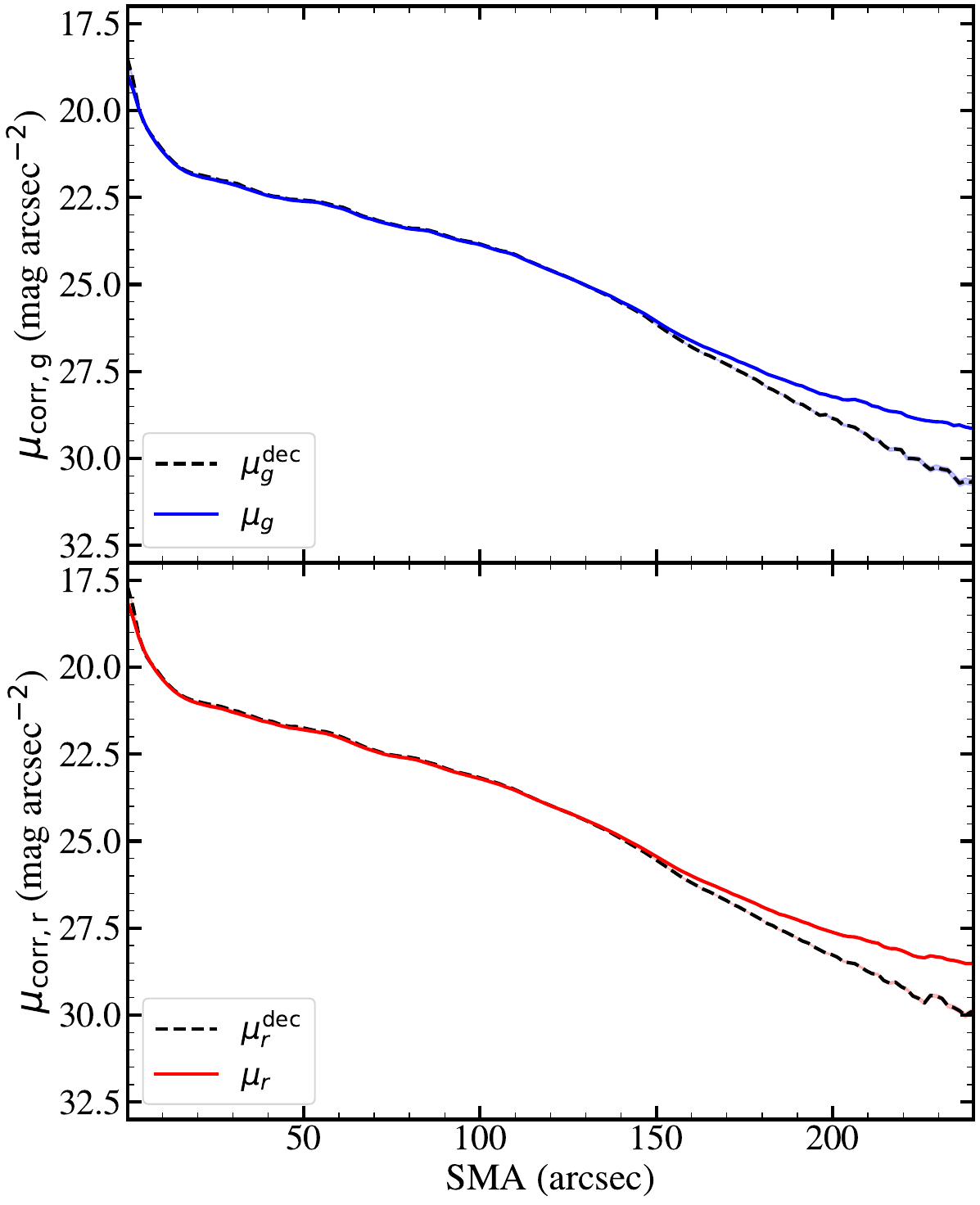}
    \caption{Effect of wavelet-based PSF deconvolution on the surface brightness profiles of NGC 7606 in the $g$- and $r$-bands. The solid curve shows the profile prior to deconvolution while the dashed curve shows the profile after deconvolution.}
    
    \label{fig:psf_deconv_comparison}
    
\end{figure}

The output of the deconvolution yields a clean, PSF-corrected image in which light redistributed by the PSF has been reconcentrated towards its physical origin, effectively restoring the intrinsic structure of the galaxy. Residual analysis shows a consistent pattern: central surface brightness increases while the outer profiles steepen, matching theoretical expectations for PSF broadening effects. This correction is essential for an accurate estimation of parameters such as the central mass surface density within 1 kpc ($\Sigma_{1,\mathrm{kpc}}$) and the $R_1$ radius, and was systematically applied to all galaxies in the sample.

To evaluate the impact of the point spread function (PSF) on our surface brightness measurements, we applied a wavelet-based deconvolution algorithm to the $g$- and $r$-band images of a representative galaxy in our sample, NGC~7606. The resulting profiles, shown in Fig.~\ref{fig:psf_deconv_comparison}, illustrate the expected redistribution of light due to PSF broadening: after deconvolution, the central regions of the galaxy become noticeably brighter, while the outer parts exhibit a steeper decline in surface brightness. This effect highlights how the PSF tends to spread central light towards the outskirts, artificially flattening observed profiles if left uncorrected. The test validates the importance of PSF treatment, particularly for the analysis of low surface brightness features.

\section{Masking the data and subtracting the local background}
\label{app:masking_data}
In the analysis of low surface brightness images, the construction of accurate masks is of paramount importance, as even faint, unmasked sources can introduce significant biases in photometric measurements. In this work, a conservative masking strategy was adopted by combining the outputs of multiple algorithms: \texttt{NoiseChisel} and \texttt{Segment} \citep{akhlaghi2015, akhlaghi2019} along with Max-Tree Objects (\texttt{MTO}; \citealt{teeninga2016,faezi2024}). The final masks were obtained by merging the individual masks produced by these methods, ensuring a robust identification of contaminant sources.

With regard to background estimation and following the methodology described in \citet{golini2025}, we accounted for possible local background variations that may not be fully captured by the global sky subtraction applied during the initial data reduction. Prior to the application of the wavelet-based deconvolution, elliptical isophotes centred on each galaxy were constructed, sharing the same position angle (PA) and axis ratio. These structural parameters were derived using the \texttt{photutils} package \citep{jedr1987, bradley2019}, where elliptical fits were performed iteratively until a satisfactory solution was reached. The resulting isophotes were visually inspected and manually adjusted in cases where the automated procedure failed to converge or yielded poor fits.

To robustly estimate the local background, we made use of wedges, which we defined as narrow, sector-shaped regions extending radially outwards from the galaxy centre. These wedges trace the galaxy light along specific angular directions and are particularly useful to study the outskirts while minimising contamination. Wedge-shaped profiles were extracted in directions carefully selected to avoid bright stars or background galaxies, allowing us to analyse local background fluctuations in well-defined sectors of the image. In each case, the sky level was estimated by computing the median pixel value within a region deemed free of contaminating flux.

\section{Additional analysis of the mass--size relation}
\label{sec:add_analysis}

In this appendix we present complementary figures supporting the analysis of the scatter in the mass--size relation. Fig.~\ref{fig:fits_mass_size} shows the linear fits to the $\log R_1$--$\log M_*$ relation for BEARD, BL, BI, and BD galaxies. The fitted slopes and intercepts for each population are listed in Table~\ref{tab:mass_size_stats}, highlighting the systematic differences between the morphological categories.  

In addition, Fig.~\ref{fig:scatter_comparison} illustrates the statistical comparison between BEARD and IllustrisTNG50 BL galaxies. In this test, $10^4$ random realisations of BL galaxies were drawn in the stellar mass range $10.0 < \log(M_*/M_\odot) < 10.5$, each time matching the number of BEARD galaxies in this interval. The resulting distribution of scatters demonstrates that the observed scatter in BEARD is fully compatible with the expectations from simulations, although Kolmogorov–Smirnov tests indicate that the two samples are statistically distinguishable in detail.
\renewcommand{\thefigure}{E.\arabic{figure}}
\setcounter{figure}{0} 
\begin{figure}
    \centering
    \includegraphics[width=\linewidth]{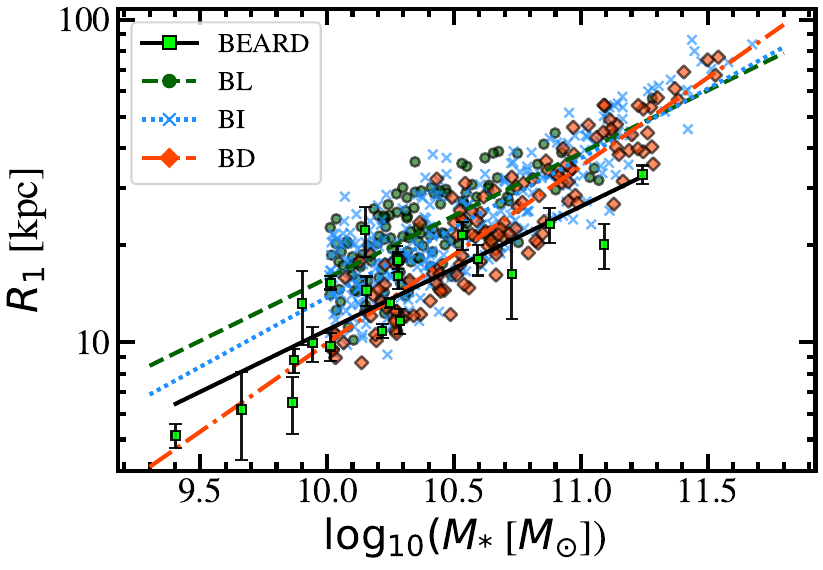}
    \caption{Linear fits to the $\log R_1$--$\log M_*$ relation for BEARD, BL, BI, and BD galaxies. Dashed lines show the best fits for each population; see Table~\ref{tab:mass_size_stats} for fitting parameters.}
    \label{fig:fits_mass_size}
\end{figure}

\begin{figure}
    \centering
    \includegraphics[width=\linewidth]{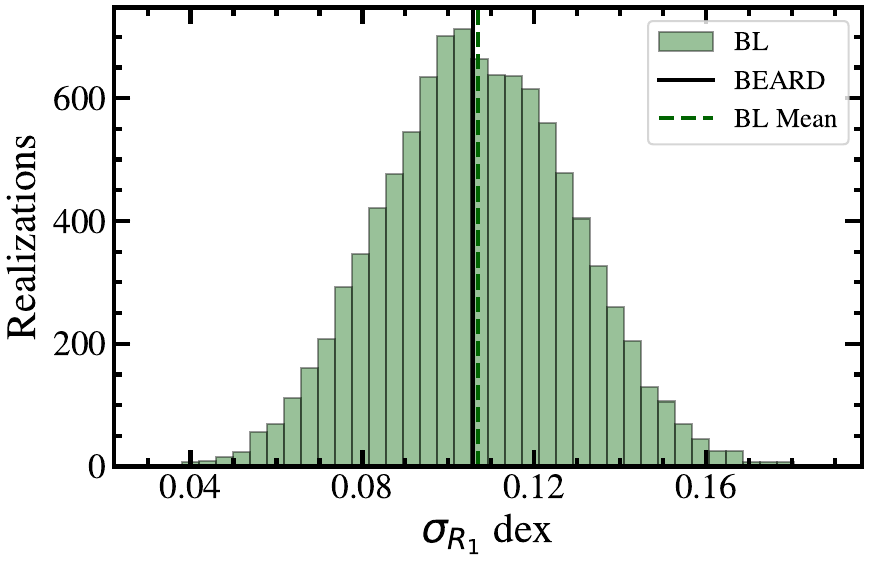}
    \caption{Distribution of scatter in $\log R_1$ for $10^4$ random realisations of IllustrisTNG50 BL galaxies in the stellar mass range $10.0 < \log(M_*/M_\odot) < 10.5$, each time matching the number of BEARD galaxies in this interval. The vertical solid line shows the BEARD scatter, while the dashed line marks the BL mean.}
    \label{fig:scatter_comparison}
\end{figure}

\renewcommand{\thefigure}{G.\arabic{figure}}
\setcounter{figure}{0} 
\begin{figure}[h]
    \centering
    \includegraphics[width=\linewidth]{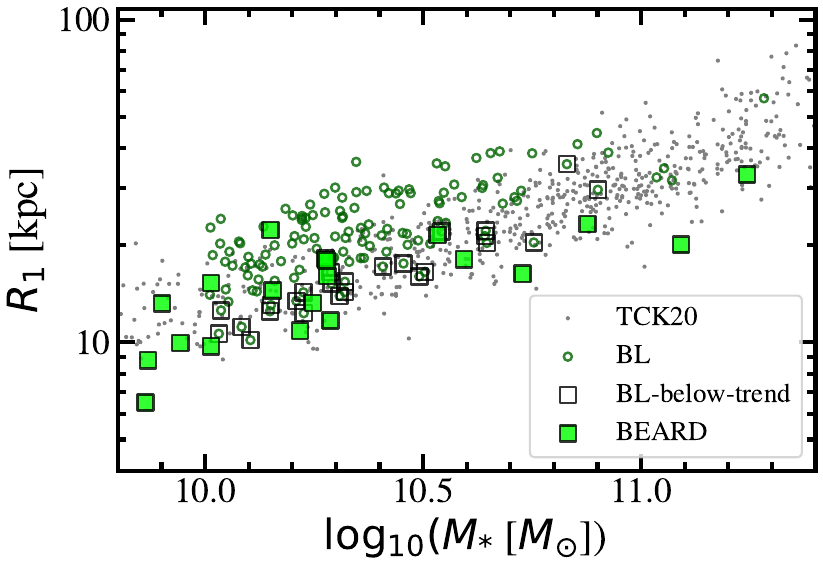}
    \caption{Stellar mass–size relation using $R_1$. The BEARD observational sample is shown in light-green squares and the TCK20 reference sample in grey dots. The population of simulated bulgeless galaxies from the IllustrisTNG50 sample is shown in green (BL), while the subsample lying below the mean mass–size trend is highlighted in black squares (BL-below-trend).}
    \label{fig:app_mass_size_bl}
\end{figure}

\section{Merger parameters and their relation to halo spin and mass}
\label{sec:mergers_spin_mass}
\renewcommand{\thefigure}{F.\arabic{figure}}
\setcounter{figure}{0} 
\begin{figure*}[t]
    \centering
    \includegraphics[width =\linewidth]{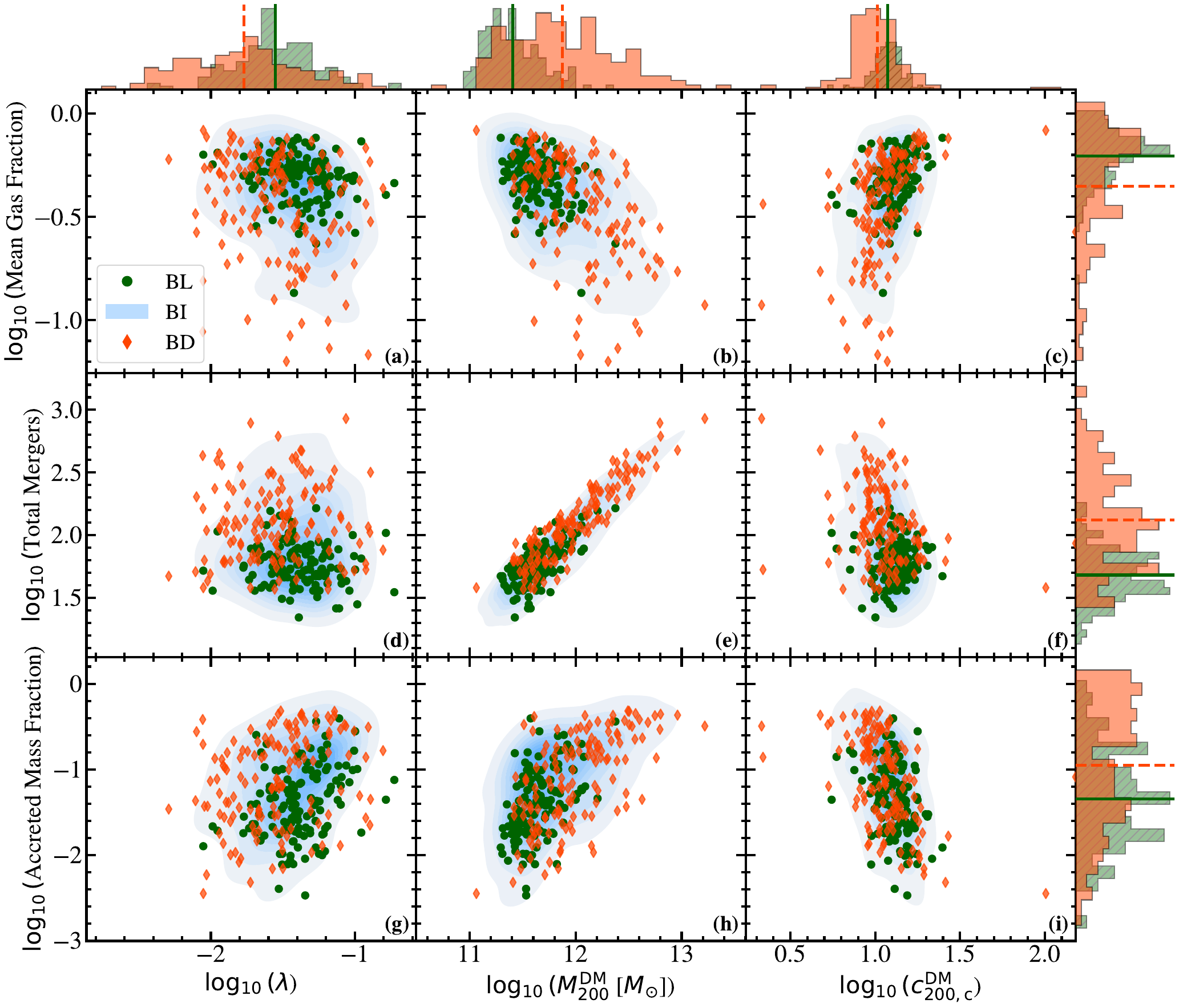}
    \caption[\footnotesize{Sigma1}]{\footnotesize{Relations between merger and dark matter properties for the simulated galaxies in our study. The figure is organised as a matrix of scatter plots, where each row corresponds to a merger property (from left to right: mean gas fraction, total mergers since $z=5$ and accreted mass fraction), and each column corresponds to a dark matter quantity (from top to bottom: spin parameter $\lambda$, halo mass $M_{200}^{\mathrm{dm}}$, and halo concentration $c_{200,\textrm{c}}^{\textrm{DM}}$). BL, BI and BD are shown as green circles, blue isodensity contours, and orange diamonds, respectively. On the top and right panels, histograms of the corresponding quantities are shown for BL and BD galaxies, where vertical lines represent the mean of each distribution.}}
    
    \label{fig:darkmatterVSmergers}
\end{figure*}

To further explore the interplay between merger history and dark matter halo properties, we analyse the correlations between three key merger parameters—total number of mergers since $z=5$, mean gas fraction, and accreted mass fraction—and three fundamental halo quantities: the spin parameter ($\lambda$), the concentration ($c_{200,\textrm{c}}^{\textrm{DM}}$) and the halo mass ($M^{\rm DM}_{200}$). The results are summarised in Fig.~\ref{fig:darkmatterVSmergers}.

Focusing first on the spin parameter, we find no strong global correlation with any of the merger-related variables. Nevertheless, the previously observed morphological separation persists: bulgeless galaxies (BL) tend to exhibit slightly higher spin values than their bulge-dominated (BD) counterparts, regardless of the merger parameter considered. This subtle separation reinforces the idea that high-spin halos are more likely to host galaxies with reduced central mass concentrations, potentially due to their merger histories being less conducive to bulge growth.

When examining the merger parameters as a function of halo mass, a noticeable positive correlation emerges between the total number of mergers and $M^{\rm DM}_{200}$. More massive halos tend to have undergone a larger number of mergers since $z=5$, consistent with expectations from hierarchical structure formation. In contrast, the remaining panels—mean gas fraction and accreted mass fraction versus halo mass, and also the ones about the concentration—do not show significant trends or morphological distinctions.

\section{Subsample of bulgeless galaxies below the mass--size trend}
\label{sec:app_bl_scatter}

In this appendix, we provide additional figures related to the analysis of a specific subsample of bulgeless galaxies from the IllustrisTNG50 simulation. As discussed in the main text, a notable fraction of bulgeless systems appears to deviate from the typical location of this morphological class in the mass--size plane, falling below the average $1 \ \sigma$ trend traced by simulated bulgeless galaxies. To investigate their properties in more detail, we isolate a subsample of such galaxies—referred to as below-trend—which lie systematically below the mean $R_1$ values of bulgeless galaxies in bins of stellar mass. Fig.~\ref{fig:app_mass_size_bl} shows the location of this below-trend population in the mass--size diagram, highlighting their offset relative to the main bulgeless sequence. 

\end{appendix}
\end{document}